\documentclass{aa} 
\usepackage{amsmath} 
\usepackage{txfonts}
\usepackage{graphicx} 
\usepackage{url}
\usepackage{natbib} 
\bibpunct{(}{)}{;}{a}{}{,}

\newcommand{\el}[2]{$\rm{}^{#2}\kern-0.6pt#1$}
\newcommand{\elm}[2]{\rm{}^{#2}\kern-0.8pt\rm{#1}}

\usepackage{color}

 \usepackage{threeparttable}

\begin{document}

\title{Effects of thermohaline instability and rotation-induced mixing on the evolution of light elements in the Galaxy : D, $^{3}$He and $^{4}$He}

\author{N. Lagarde\inst{1} \and D. Romano\inst{2} \and C. Charbonnel\inst{1,3} \and M. Tosi\inst{2} \and C. Chiappini\inst{4} \and F. Matteucci\inst{5,6}} 

\offprints{N. Lagarde,\\ email: Nadege.Lagarde@unige.ch}

\institute{Geneva Observatory, University of Geneva, Chemin des
  Maillettes 51, 1290 Versoix, Switzerland \and
              INAF-Bologna Observatory, Via Ranzani 1, I-40127, Bologna, Italy\and
              IRAP, CNRS UMR 5277, Universit\'e de Toulouse, 14, Av. E.Belin, 31400 Toulouse, France\and
              Leibniz-Institut f\"ur Astrophysik Potsdam (AIP), An der Sternwarte 16 D-14482, Potsdam, Germany\and
              Physics Department, Trieste University, Via Tiepolo 11, I-34143 Trieste, Italy\and
              INAF-Trieste Observatory, Via Tiepolo 11, I-34143 Trieste, Italy\
}

    \date{Submitted 29 February 2012/ Accepted 8 April 2012}

\authorrunning{N. Lagarde et al.}  \titlerunning{Evolution of light elements in the Galaxy : D, $^{3}$He and $^{4}$He}

\abstract
   { Recent studies of low- and intermediate-mass stars show that the evolution of the chemical elements in these stars is very different from that proposed by standard stellar models. Rotation-induced mixing modifies the internal chemical structure of main sequence stars, although its signatures are revealed only later in the evolution when the
first dredge-up occurs. Thermohaline mixing is likely the dominating process that governs the photospheric composition of low-mass red giant branch stars and has been shown to drastically reduce the net $^{3}$He production in these stars.
The predictions of these new stellar models need to be tested against galaxy evolution. In particular, the resulting  evolution of the light elements D, $^{3}$He and $^{4}$He should be compared with their primordial values inferred from the Wilkinson Microwave Anisotropy Probe data and with the abundances derived from observations of different Galactic regions.
  } 
   {We study the effects of thermohaline mixing and rotation-induced mixing on the evolution of the light elements in the Milky Way.}
   {We compute Galactic evolutionary models including new yields from stellar models computed with thermohaline instability and rotation-induced mixing. We discuss the effects of these important physical processes acting in stars on the evolution of the light elements D, $^{3}$He, and $^{4}$He in the Galaxy.}
   {Galactic chemical evolution models computed with stellar yields including thermohaline mixing and rotation fit better observations of $^{3}$He and $^{4}$He in the Galaxy than models computed with standard stellar yields.}
   {The inclusion of thermohaline mixing in stellar models provides a solution to the long-standing ``$^{3}$He problem'' on a Galactic scale. Stellar models including rotation-induced mixing and thermohaline instability reproduce also the observations of D and $^{4}$He.}


\maketitle

\section{Introduction}
\label{section:introduction}

Understanding the evolution of the light elements deuterium (D), helium-3 
($^3$He), and helium-4 ($^4$He) hinges on the comprehension of several 
astrophysical processes and links together different branches of physics and 
cosmology. D, $^3$He, and $^4$He are all synthesized in astrophysically 
relevant  quantities during Big Bang Nucleosynthesis \citep[BBN;][]{Peebles66,
Wagoner67}. In the absence of any other realistic production channel 
\citep{Epstein76,Prodanovic03}, the abundance of D in galaxies smoothly 
decreases in time as gas cycles through stars. $^3$He and $^4$He, instead, have 
a more complex history, since they are both produced and destroyed in stars.

The sensitivity of the predicted BBN abundance of D, (D/H)$_{\mathrm{P}}$, to 
the baryon-to-photon ratio, $\eta_{\mathrm{B}}$, coupled with its 
straightforward galactic evolution, has long made this element be the 
`baryometer' of choice \citep{Reeves73}. In the nineties, the controversial 
assessment of the primordial abundance of D from observations of high-redshift 
clouds \citep[e.g.][]{Songaila94,BuTy98}, probing an almost unevolved medium, 
led researchers to resort to detailed Galactic chemical evolution (GCE) 
modeling in order to constrain the pre-Galactic D abundance and, hence, the 
value of $\eta_{\mathrm{B}}$. It was shown that any reasonable GCE model results 
in moderate local D astration, by less than a factor of three 
\citep{Steigman92,Edmunds94,Gallietal95,Dearborn96,Prantzos96,Tosi98,
Chiappini02}. Assuming that the local present-day abundance of D is well known, 
this provided a stringent bound to (D/H)$_{\mathrm{P}}$ and, hence, a test for 
BBN theories. Chemical evolution models able to reproduce the majority of the
observational constraints for the Milky Way firmly ruled out the highest
values of (D/H)$_{\mathrm{P}}$ by \citet{Songaila94}. In particular, the values 
of (D/H)$_{\mathrm{P}}$ and $\eta_{\mathrm{B}}$ suggested by \citet{Tosi00} and \citet{Chiappini02} 
turned out to be in very good agreement with the ones determined from the 
analysis of the first \emph{Wilkinson Microwave Anisotropy Probe} (\emph{WMAP}) 
data \citep{Bennett03,Spergel03} .

\begin{table*}[t]
  \caption{Abundances of Deuterium, $^{3}$He, and $^{4}$He at different epochs. }
  \centering
   \begin{threeparttable}
  \begin{tabular}{c c c c c c }
    \hline
    Chemicals & Units &SBBN+\emph{WMAP} & Protosolar cloud & Local intestellar medium \\ 
                        & &  (t=0 Gyr)      &  (t=9.2 Gyrs )       &(Present time)\\  
    \hline
    \hline
    	D	         &	10$^{5}$ (D/H)	        & 	2.49 $\pm$ 0.17$^{a}$          &     2.1 $\pm$ 0.5$^{c}$   &   2.31$\pm$ 0.24$^{d}$ \\
	                 &                                             & 2.60 $^{+ 0.19}_{- 0.17}$$^{b}$ &                                  & 0.98 $\pm$ 0.19$^{e}$ \\
	                 &                                            &                                                &                                 & 2.0 $\pm$ 0.1$^{f}$\\
	                 \hline
    	$^{3}$He	& 10$^{5}$ ($^{3}$He/H) &	1.00 $\pm$0.07 $^{a}$            &      1.5 $\pm$ 0.2$^{c}$  &   2.4 $\pm0.7^{g}$ \\
	                   &                                          &      1.04 $\pm$ 0.04$^{b}$             &                               & \\
	                   \hline
    $^{4}$He	& Y	                                    &	0.2486 $\pm$ 0.0002$^{a}$  &	 0.2703$^{h}$                &                            \\
                            &                                           &0.2479 $^{+ 0.0004}_{- 0.0005}$$^{b}$&                                 & \\ 
    \hline
  \end{tabular}
  \label{tab:obs}
      \begin{tablenotes}
       \item (a) \cite{Cyburt08}; (b) \citet{Coc04};  (c) \citet{GeGl98} ; (d) \citet{Linsky06} ;  (e) \citet{Hebrard05} ;  (f) \citet{Prodanovic10} ;   (g) \citet{GlGe96} ;  (h) \citet{Asplund09} 
	\end{tablenotes}
	\end{threeparttable}
\end{table*}

Indeed, the first release of results from \emph{WMAP} made us enter a new 
precision era for cosmology -- 
$\eta_{\mathrm{B}}$ is now known with exquisite accuracy. Converging 
measurements of D abundances in remote gas clouds lead to (D/H)$_{\mathrm{P}}$~= 
(2.8 $\pm$ 0.2)~$\times$ 10$^{-5}$ \citep{Pettini08}, a value consistent, 
within the errors, with the primordial D abundance predicted by the standard 
model of cosmology with parameters fixed by \emph{WMAP} data \citep[e.g.][see 
Table~\ref{tab:obs}]{Cyburt08,Coc04}. The remarkable homogeneity of D 
abundances at high redshifts, however, clashes with the unexpectedly large 
scatter in D/H revealed by determinations of relatively local D 
abundances \citep{Vidal-Madjar98,Jenkins99,Sonneborn00,Hebrard02,Oliveira06}. 
The observed dispersion can be reconciled with the predictions on D evolution 
from standard GCE models by taking into account two short-term, small-scale 
phenomena: D depletion onto dust grains and localized infall of primordial gas 
\citep{Romano06,Steigmanetal07,Romano10}.

As far as $^3$He is concerned, \citet{Iben67} and \citet{TruCa71} first showed 
that large amounts of this element are produced by low-mass stars ($M \simeq$ 
1--3 M$_{\odot}$) in the ashes of hydrogen burning by the $p$-$p$ cycle on the 
main sequence. Problems arised soon, when \citet{Roodetal76} incorporated these 
yields in models of GCE and came out with predicted $^3$He abundances orders of 
magnitude higher than the observed ones in the Galaxy. Indeed, the nearly 
constancy of the $^3$He abundance with both time and position within the Galaxy 
\citep[e.g.][]{Bania02} rather implies a negligible production of this element 
in stars, at variance with predictions from standard stellar models. It was 
then advocated that some non-standard mechanisms are acting in a major
fraction of low-mass stars, which prevents the fresh $^3$He from surviving and 
being ejected in the interstellar medium (ISM; \citealt{Roodetal1984}). 
For instance, extra mixing was also invoked in order to explain other abundance anomalies like the carbon isotopic ratio in red giants (\citealt{Charbonnel95,Hogan1995,CharDoNa98}; see also \citealt{Eggleton06,Eggleton08}).
The first  $^{3}$He stellar yields for low-mass stars computed with ``ad hoc'' extra-mixing (i.e., not related to a physical mechanism, see Section 2.2.2) became available with the work of \citet{BoSa99}. \citet{Galli97,Palla00,Chiappini02} implemented those yields, accounting for the dependency of the helium-3 production/destruction on the stellar mass, metallicity and initial D abundances. It was shown that chemical evolution models which account for about 90\% of low-mass stars undergoing extra-mixing led to a good agreement with the Proto Solar Cloud (PSC) observations as well as with the observed gradient along the disk.

The physical process responsible for the extra-mixing has possibly been recently 
identified. \cite{ChaZah07a} have shown indeed that thermohaline instability,  when modelled with a simple prescription based on linear stability analysis, leads to 
a drastic reduction of $^3$He production in low-mass, low-metallicity red giant stars. However a couple of planetary nebulae, namely NGC 3242 and J320, have been found
to behave ÒclassicallyÓ \citep[see][]{Bania10}: slightly more massive than the Sun, they
are currently returning fresh $^3$He to the ISM, in agreement with standard predictions \citep{Rood92, Galli97, Balser99a, Balser06}.
To reconcile the $^3$He/H measurements in Galactic HII regions with the high values of
$^3$He in NGC 3242 and J320, \citet{ChaZah07b} proposed that thermohaline
mixing is inhibited by a fossil magnetic field in red giant branch (RGB) stars that are descendants of Ap
stars. The percentage of such stars is about 2-10\% of all A-type objects. This number agrees with the fact that about 4\% of low-mass  evolved stars (including NGC 3242) exhibit standard suface abundances as depicted by their carbon isotopic ratios \citep{CharDoNa98}.
Grids of $^3$He yields from stellar models taking into account thermohaline 
instability, as well as rotational mixing, are presented 
in \cite{Lagarde11}. These are suitable for use in GCE studies such as the one presented here.

The scarce determinations of reliable $^4$He abundances in the Galaxy 
\citep{Balser10,Peimbert10}, together with the steadiness of GCE model predictions on the evolution of $^4$He when adopting different stellar yields \citep[e.g.][]{Chiappini02,RKTM10}, have given this element little attention in GCE studies.

In this paper, we deal with the evolution of D, $^3$He, and $^4$He in the solar 
vicinity, as well as their distributions across the Galactic disc. The 
production of $^3$He is strictly related to the destruction of D and the 
production of $^4$He. Therefore, these elements are considered all together. 
The main novelty of the present work is the new nucleosynthesis prescriptions for the synthesis of $^{3}$He and $^{4}$He in low-mass stars (masses below 6 M$_{\odot}$). Here we present, for the first time, chemical evolution models for $^{3}$He computed with stellar yields from non-standard stellar models that include both rotation-induced and thermohaline mixing \citep{ChaZah07b,ChaLag10,Lagarde11}.  As we will show in the next sections, these yields  bring chemical evolution model predictions into agreement with the Galactic $^{3}$He data without the necessity of assuming "ad hoc" fractions of extra-mixing among low-mass stars.

The layout of the paper is as follows.  
In Section ~\ref{nucleosynthesis} we discuss the production/destruction channels of D, 
$^3$He, and $^4$He in stars in light of new generation stellar models that take into account rotation-induced mixing and thermohaline instability. A comparison with stellar yields predictions from the literature is presented. In Section~\ref{chemcode}  we describe the adopted GCE model. 
In Section~\ref{lightelementevol} we discuss our results on the evolution of 
the light elements in the Galaxy. In particular, we try to put constraints on 
the uncertain parameters of stellar evolution through a comparison of the model 
predictions with the relevant data. 
We draw our conclusions in Section~\ref{conclusions}.

\section{Stellar nucleosynthesis of light elements 
from new generation models 
for low- and intermediate-mass stars}
\label{nucleosynthesis}

In a series of papers (\citealt{ChaLag10,Lagarde11, Lagarde12a}, hereinafter Papers I, II and III, respectively; see also \citealt{ChaZah07a,ChaZah07b}), we discuss the impact of rotation-induced mixing and thermohaline instability on the structure, evolution, nucleosynthesis and yields, as well as on the asteroseismic and chemical properties of low- and intermediate-mass stars at various metallicities. A detailed description of the input physics of the models is given in Paper III. Rotation-induced mixing is treated using the complete formalism
developed by \citet{Zahn92} and \citet{MaeZah98} (see for more details, Papers I, II, III).
In these stellar models, we consider that thermohaline instability develops as long thin fingers with the aspect ratio consistent with predictions by \citet{Ulrich72} and confirmed by the laboratory experiments \citep{Krish03}. We note that this value is higher than obtained by current 2D and 3D numerical simulations  \citep{Denissenkov10, DenissenkovMerryfield10, RosenblumGaraudetal11, Traxler11}. Before a final word on this discrepancy comes from future numerical simulations in realistic stellar conditions, we adopt an aspect ration (i.e., maximum length relative to their diameter) of 5, which nicely accounts for the observed chemical properties of red giant stars.
These new generation stellar models account very nicely for main-sequence and RGB abundance patterns observed in field and open cluster stars over the mass and metallicity range covered, as shown in Paper I and  in \cite{ChaZah07a}. In this section we briefly summarize their characteristics as far as the nucleosynthesis of light elements is concerned.

\subsection{Deuterium}

Deuterium 
burns by proton-captures at low temperature 
during the pre-main sequence \citep{Reeves73}. 
Therefore, it is totally destroyed in stellar interiors and the corresponding net yields are negative whatever the mass and metallicity of the star, as in the case of standard stellar models.

\subsection{Helium-3} \label{He3_BS}
\subsubsection{Impact of rotation-induced mixing and thermohaline instability on model predictions}
\label{3hestellarmodels}

\begin{figure*}
	\centering
      \includegraphics[angle=0,width=9cm]{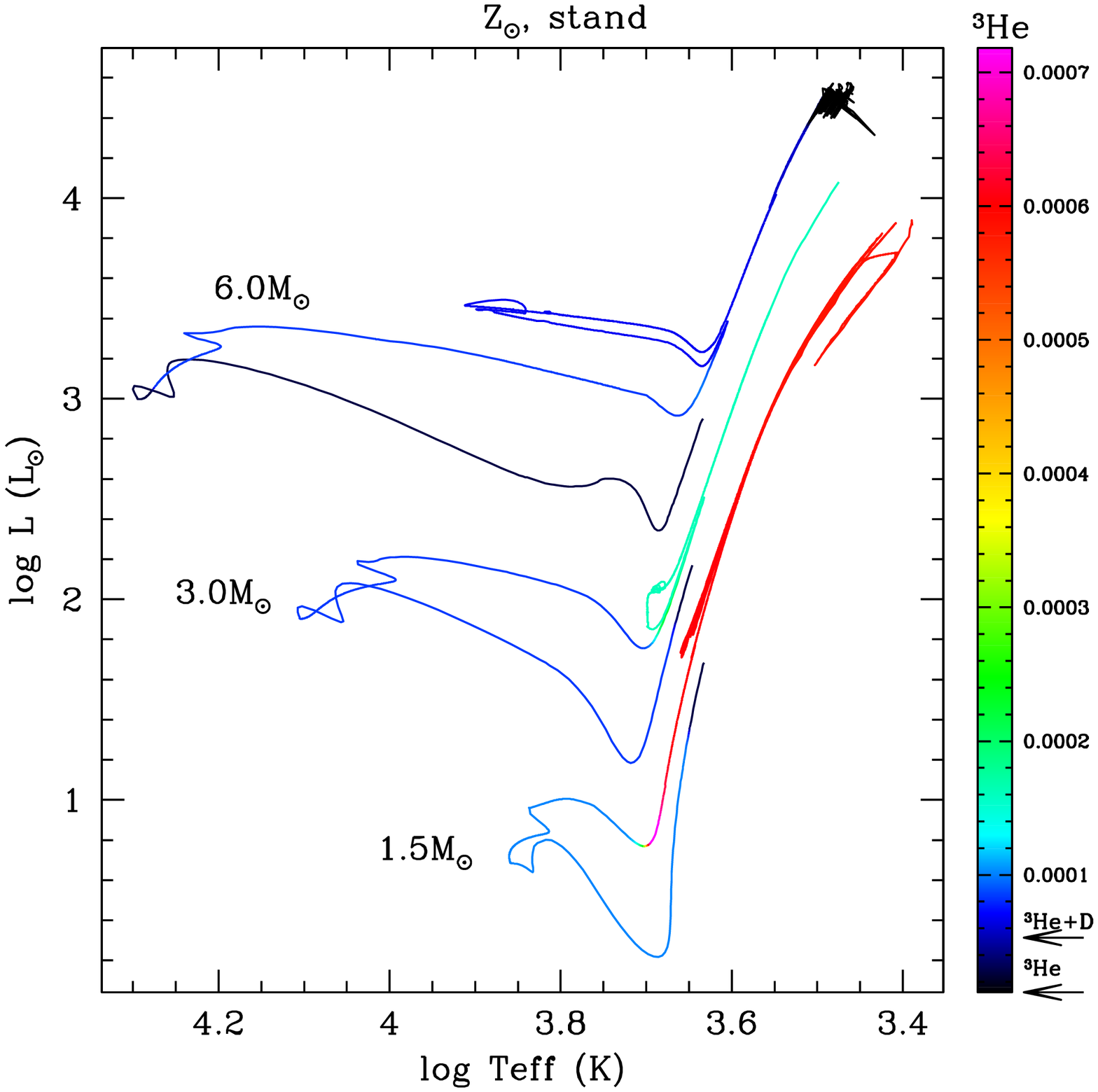}
  \includegraphics[angle=0,width=9cm]{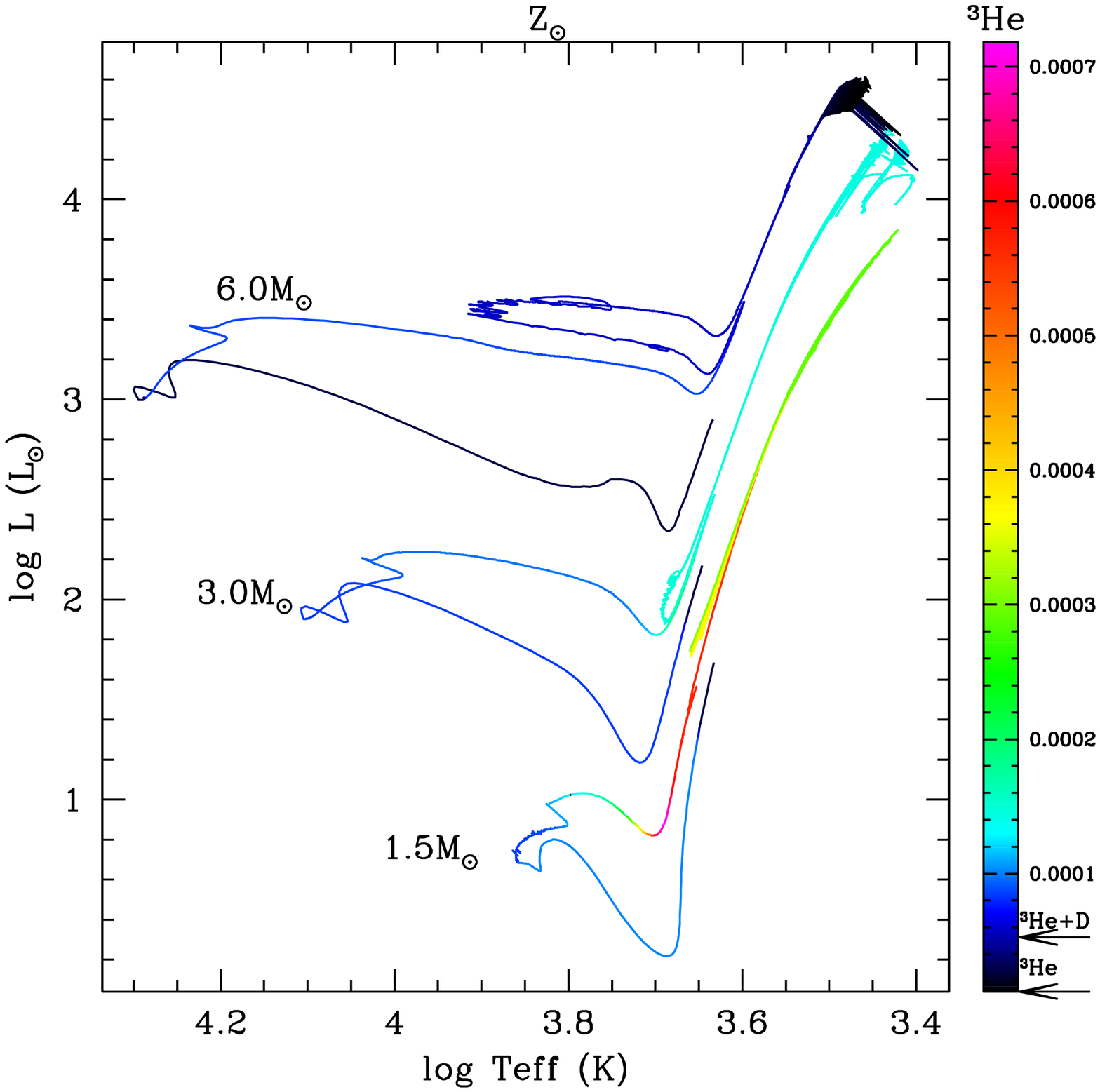}
   \caption{Theoretical evolutionary tracks in the HR diagram for the 1.5~M$_{\odot}$, 3.0~M$_{\odot}$, and 6.0~M$_{\odot}$ models at solar metallicity 
following the standard prescription (left panel) ; and  including rotation-induced mixing and thermohaline instability (right panel), from the pre-main sequence up to the end of the TP-AGB phase.
Colours depict the mass fraction of $^{3}$He at the
stellar surface as indicated on the right, with the arrows showing the initial $^{3}$He and $^{3}$He$+^{2}$H content assumed at birth.}
   \label{fig:Hrd_He3}
\end{figure*}

\begin{figure*} 
	\centering
		\includegraphics[angle=0,width=9cm]{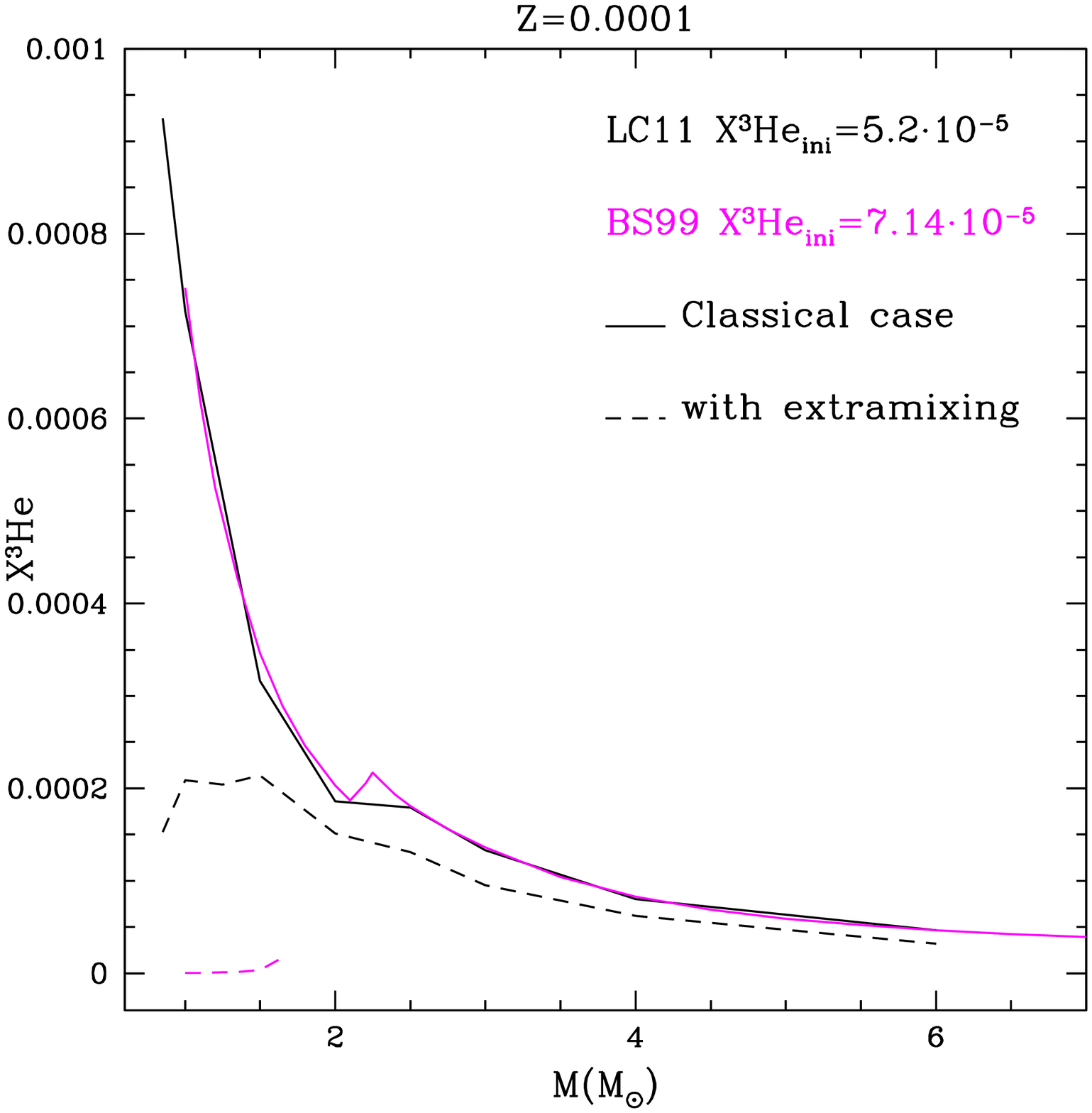}
		\includegraphics[angle=0,width=9cm]{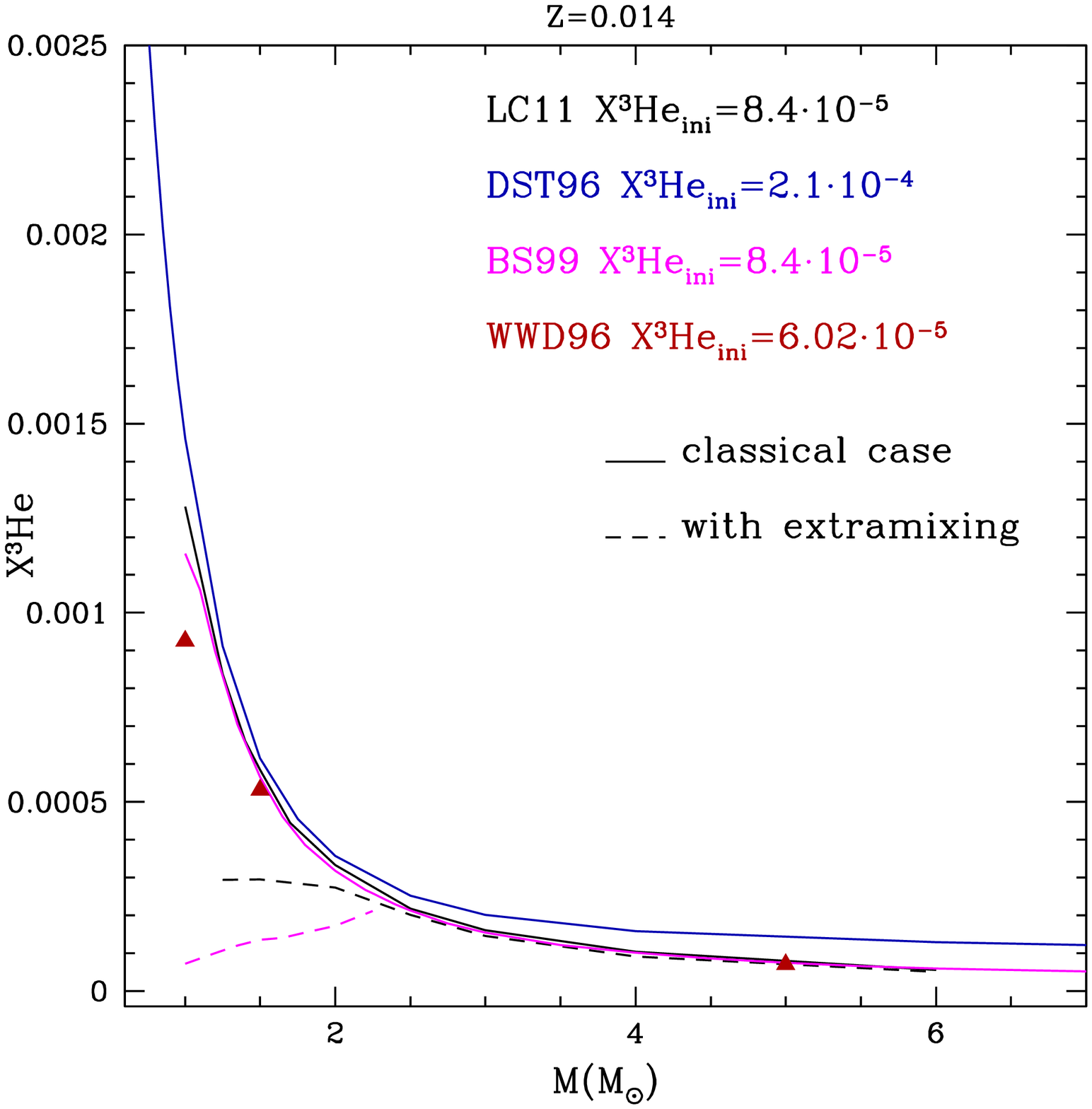}
	  \caption{Mass fraction of $^{3}$He at the stellar surface
at the end of the second dredge-up as a function of initial stellar mass 
in standard models (solid lines and triangles) and in models including 
various prescriptions for mixing in the radiative regions (dashed lines), at two metallicities (Z=0.0001 and Z=0.014; left and right panels, respectively). 
The predictions are from \citet[DST96, blue line]{Dearborn96}, \citet[BS99, at the RGB tip, magenta lines]{BoSa99}, \citet[WWD96, red triangles]{Weissetal96}, and \citet[][LC11, black lines]{Lagarde11}.
The initial abundances of $^{3}$He adopted at the zero age main sequence by 
the different authors are given on the top right corner of each panel.}
	\label{fig:XHe3}
\end{figure*}

Paper II describes in detail the behaviour of $^3$He both in the standard case and in our models including thermohaline instability and rotation-induced mixing. 
Figure~\ref{fig:Hrd_He3} shows the theoretical evolution of the $^{3}$He surface abundance along the 
evolutionary tracks in the HR diagram for the models of 1.5, 3.0, and 6.0~M$_{\odot}$ at solar metallicity that take into account these two processes. 
The colour coding in mass fraction is given on the right, with the initial values of $^{3}$He and D$+$$^{3}$He indicated by the arrows. Evolution is shown from the pre-main sequence along the Hayashi track up to the end of the thermally-pulsing asymptotic giant branch (TP-AGB) phase.
We see the changes in $^{3}$He surface abundance due to D-burning on the pre-main sequence except in the 6.0~M$_{\odot}$ model whose convective envelope withdraws very quickly at the beginning of that phase allowing to the preservation of pristine D in a very thin external layer. 
$^{3}$He is then produced in low- and intermediate-mass stars during the main sequence through the pp-chains and subsequently dredged-up when the stars move towards the RGB. 
On the 1.5~M$_{\odot}$ track 
one sees the effect of rotation-induced mixing that already brings fresh $^{3}$He towards the stellar surface while the star is on the main sequence (on the contrary, in the standard case the surface abundance of $^{3}$He remains constant on the main sequence and starts changing only at the very base of the RGB due to the first dredge-up). 
However, as discussed in Paper II rotation-induced mixing is found to lower the total $^{3}$He production compared to the standard case over the whole mass and metallicity range scrutinized, and to decrease the upper mass limit at which stars destroy this element (see Fig.~5 in Paper II). 
Additionally, for low-mass stars (M $\leq$ 2--2.2~M$_{\odot}$) thermohaline mixing occurring during the RGB phase beyond the bump and on the TP-AGB leads to the destruction of part of the freshly produced $^3$He while accounting for the observed surface abundance anomalies of other chemicals (i.e., lithium and nitrogen, as well as carbon isotopic ratio; see also \citealt{ChaZah07a}); 
the associated decrease of surface $^{3}$He is clearly seen for the 1.5~M$_{\odot}$ model in Fig.\ref{fig:Hrd_He3} (right panel). 
Therefore, although low-mass stars remain net  $^3$He producers, their contribution to the Galactic evolution of this element is much lower than in the standard case. It was also shown in Paper II that thermohaline mixing leads to  $^3$He depletion during the TP-AGB phase for stars with masses up to $\sim$ 4~M$_{\odot}$ (see the 3~M$_{\odot}$ track in the figure). In more massive intermediate-mass stars, $^3$He is further destroyed through hot-bottom burning on the TP-AGB. Finally, this figure shows that the lower the mass of the star, the higher the surface abundance of $^{3}$He at the end of the TP-AGB. 
The global impact of rotation-induced mixing and thermohaline instability on the net $^3$He yields from low- and intermediate-mass stars of various metallicities is summarized in Fig.9 and in Tables 1 to 4 of Paper II (see also Fig.~\ref{fig:XHe3} discussed below).

\subsubsection{Comparison with other stellar models}
\label{3Hecomparison}

In Fig.~\ref{fig:XHe3} we show the mass fraction of $^{3}$He at the end of the second dredge-up 
as a function of the initial stellar mass at two metallicities (Z=0.0001 and  0.014;  left and right panels, respectively) 
from our models  (black lines; see Paper II). A comparison is made with model predictions from the literature (coloured lines and filled triangles). 

Our standard predictions (full lines) are in very good agreement with those from \citet{Dearborn96}, \citet{BoSa99}, and \citet{Weissetal96}. For the reasons given in Section~\ref{3hestellarmodels} they are higher than in the case including rotation-induced mixing and thermohaline instability (dashed black lines). Let us note that thermohaline mixing on the TP-AGB leads to further decrease of the $^{3}$He mass fraction at the stellar surface (reduction of 64, 83, and 19\% in the [1.25~M$_{\odot}$, Z$_{\odot}$], [0.85~M$_{\odot}$, Z=0.0001], and  [2.0~M$_{\odot}$, Z$_{\odot}$ and Z=0.0001] models, respectively; see Paper II).

In Fig.~\ref{fig:XHe3} we also plot the predictions at the end of the second dredge-up for the models of \citet[][hereafter BS99]{BoSa99} that include parametric  mixing below the base of the convective envelope of RGB stars (the so-called ``conveyor-belt'' circulation and the associated ``cool bottom processing'', CBP; dashed magenta lines). In BS99 post-processing computations, the mixing is not related to any physical mechanism. Rather, the depth of the mixed zone is a free parameter that corresponds  to  the difference $\Delta$logT between the temperature at the base of the hydrogen-burning shell 
and that at the base of the assumed mixed zone.  
BS99 assumed that the value of $\Delta$logT remains constant along the RGB, and this free parameter was calibrated in order to reproduce the observed carbon isotopic ratio of clump stars in M67. 
We note that the fixed $\Delta$logT value used by 
BS99 leads to shallower mixing just after the RGB bump than what we get when considering thermohaline instability as the physical mixing mechanism.  As a consequence BS99 prescription leads to a slow and gradual theoretical decrease of the $^{12}$C/$^{13}$C ratio up to the RGB tip; this behaviour disagrees with the sudden drop of this quantity that is observed after the RGB bump and that is well reproduced by our models that include thermohaline mixing (see Paper I  and \citealt{ChaZah07a}). 
However, due to the higher compactness of the hydrogen-burning shell when the stars reach higher luminosities on the RGB, the fixed $\Delta$logT value adopted by BS99 leads to deeper mixing close to the RGB tip. 
The second free parameter is the stream mass flow rate, $\dot{M}_{p}$. 
Contrary to thermohaline mixing, for which the stream mass flow rate\footnote{To translate the thermohaline diffusion coefficient of our models into a stream mass flow rate as used by BS99, we compute : $$ \dot{M}_{p}=\frac{4\pi r^{2} \rho \mathcal{D}}{l_{mix}}$$ with $r$ and $\rho$ the radius and the density at the base of the region where the thermohaline instability develops, $l_{mix}$ the size of this mixing zone, and $\mathcal{D}$ the mean diffusion coefficient along the mixing zone.} decreases along the RGB from $\sim$10$^{-2.5}$ to $\sim$10$^{-5}$ M$_{\odot}$ yr$^{-1}$, BS99 assumed that the value of $\dot{M}_{p}$ stays constant along the RGB ($\dot{M}_{p}=10^{-4} \, \mathrm{M_{\odot} \, yr^{-1}}$). This difference in the mixing efficiency at the end of the RGB explains the difference between BS99's models and ours on the final $^{3}$He surface mass fraction. Another drawback of such models was also that a physical reason was missing to explain why only 7\% of the low mass stars would produce $^{3}$He following the standard predictions. 

\subsection{Helium-4}

\subsubsection{Standard predictions}

\begin{figure} 
	\centering
	\includegraphics[angle=0,width=9cm]{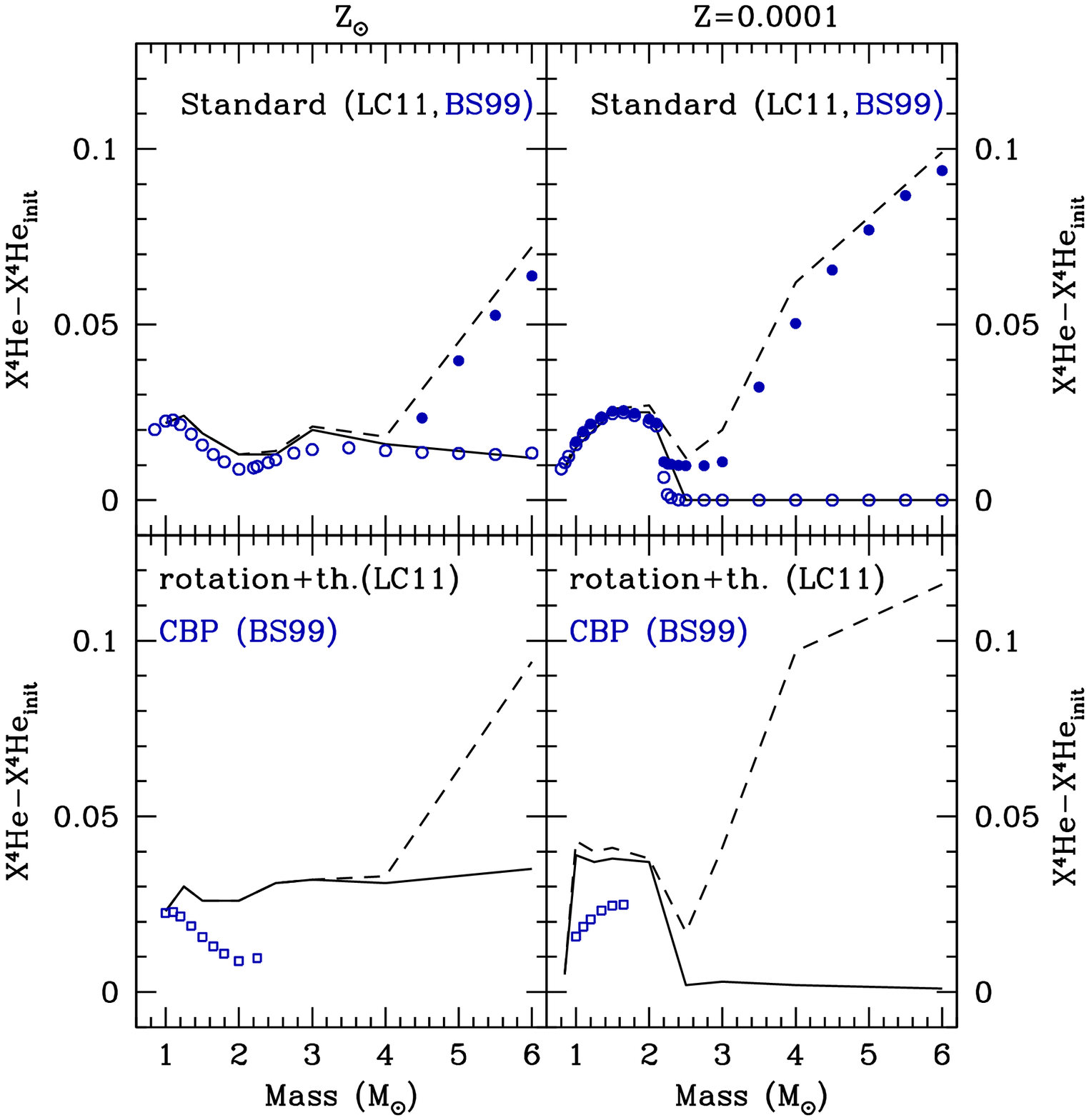}
	  \caption{Relative enrichment of the surface abundance of $^{4}$He with respect to its initial value (in mass fraction) for Paper I models at the end of the first and second dredge-up  (solid and dashed black lines, respectively) as a function of the initial stellar mass and for two metallicities (Z$_{\odot}$ and Z=0.0001; left and right panels, respectively). \textbf{\textit{Top panels:}} our standard predictions are compared with those of BS99 (blue circles; open and full at the end of the first and second dredge-up, respectively). \textbf{\textit{Bottom panels:}} our models including both rotation-induced mixing and thermohaline instability are compared with the ``cool-bottom processing" predictions by BS99 at the RGB tip (blue squares).}
	\label{fig:PXHe4_BS99}
\end{figure}

\begin{figure} 
	\centering
	\includegraphics[angle=0,width=9cm]{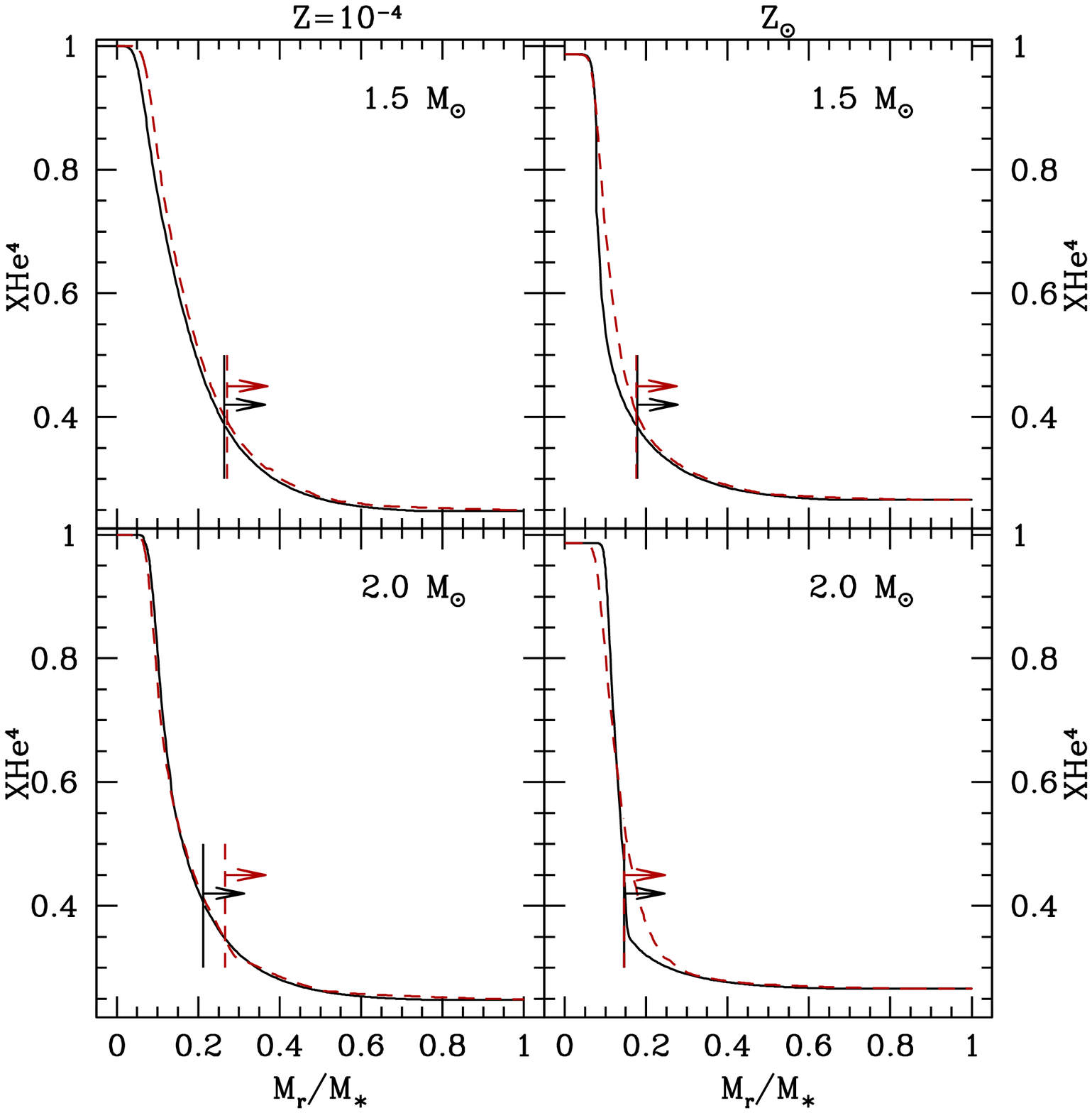}
	  \caption{Abundance profile of $^{4}$He at the end of the main sequence for the 1.5~M$_{\odot}$ and 2.0~M$_{\odot}$ models at two metallicities (Z=0.0001 and Z=0.014, left and right panel, respectively) in the standard case (solid black line), and when including the effects of rotation (red dashed line). Models are from Paper~II. The maximum depth reached by the convective envelope during the first dredge-up is shown with the vertical bars and arrows.}
	\label{fig:PXHe4}
\end{figure}

Changes in the surface $^{4}$He abundance result from the dredge-up episodes undergone by the stars along their evolution, that eventually lead to the partial engulfment of the ashes of central- and shell-H-burning by the convective envelopes. The efficiency of the successive dredge-up events depends both on stellar mass and metallicity (see e.g. Paper II), as depicted in Fig. \ref{fig:PXHe4_BS99} where we present our standard predictions (upper panels) for the surface abundance of $^{4}$He, adjusted to its initial value, in mass fraction, after the first and the second dredge-up (solid and dashed lines, respectively) as a function of the initial stellar mass, for our two extreme metallicities (Z$_{\odot}$ and Z=0.0001).
For solar metallicity, the surface abundance of $^{4}$He increases during the first dredge-up 
over the whole mass range investigated, while for the lowest metallicity (Z=10$^{-4}$) only stars with M~$\leq$ 2.5~M$_{\odot}$ are affected. On the other hand, the second dredge-up leads to an increase of the $^{4}$He surface abundance only for the upper part of the considered mass range (i.e., for stars with masses higher than $\sim$ 2.5~M$_{\odot}$).
Finally, the third dredge-up, which occurs in stars with mass $\geq$ 6.0 M$_{\odot}$ for Z$_{\odot}$ and 2.5 M$_{\odot}$ for Z=10$^{-4}$
during the TP-AGB phase, allows to increase the $^{4}$He abundance at the surface of these stars (not shown here). This increase is $\sim$ 1\% for [6.0 M$_{\odot}$, Z$_{\odot}$], and [4.0 M$_{\odot}$, Z=0.0001] models.\footnote{Note that we did not include additional mechanisms to force the third dredge-up in our models, as required in the literature for reproducing the carbon-star luminosity function \citep[e.g.][]{Frogel90,CoFr96,Groenewegen98,Marigo99}. Hence, we do not attempt any comparison of the model predictions with available data for C and N isotopes. However this has no impact on the galactic chemical evolution of the light elements we focus on.}

\subsubsection{Impact of rotation-induced mixing and thermohaline instability on model predictions}
\label{4hestellarmodels}

\begin{figure*}
	\centering
   \includegraphics[angle=0,width=9cm]{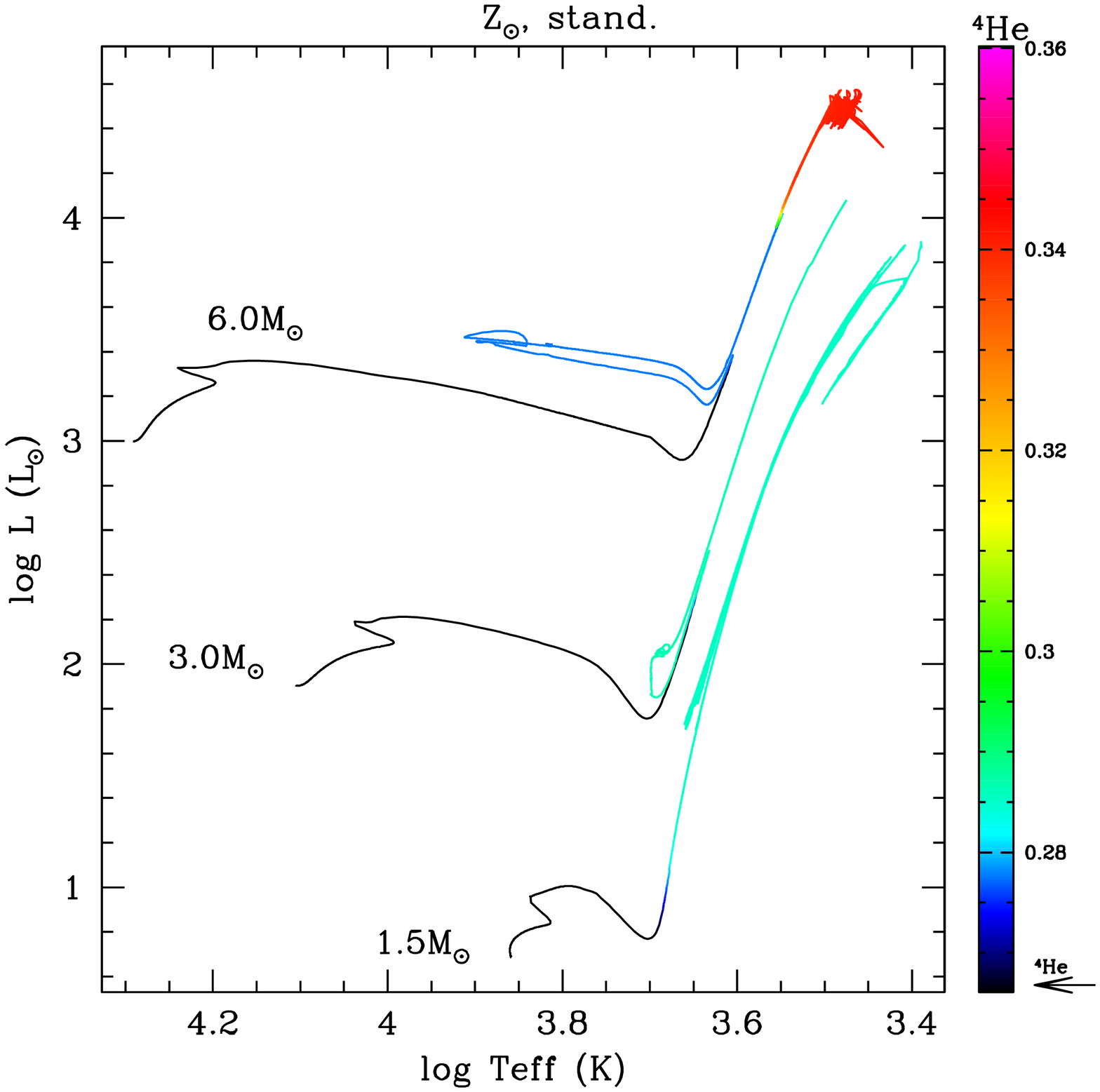}
  \includegraphics[angle=0,width=9cm]{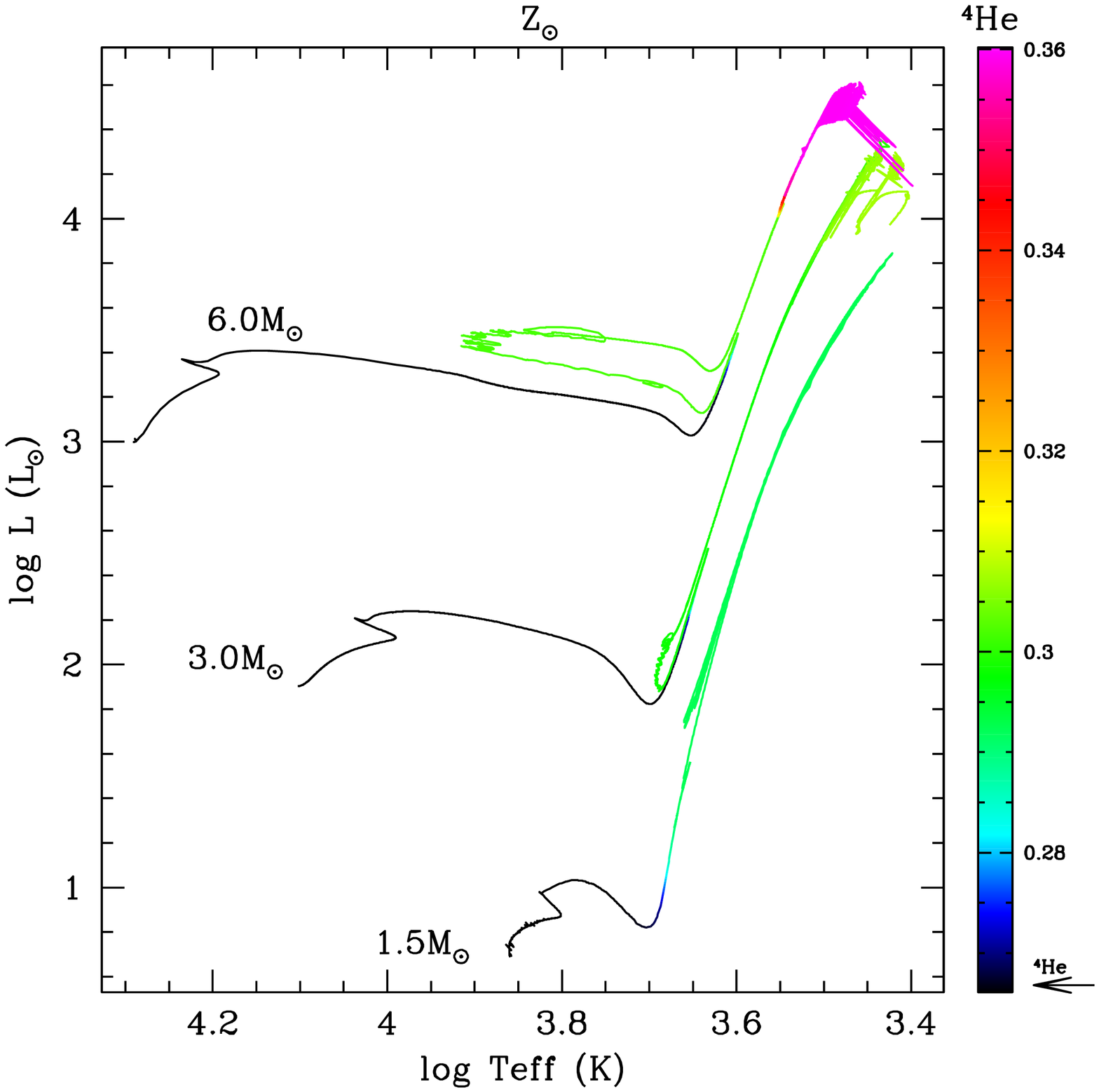}
   \caption{Same as Fig.~\ref{fig:Hrd_He3} for $^4$He from the zero age main sequence to the end of the TP-AGB.}   \label{fig:Hrd_He4}
\end{figure*}

Thermohaline instability has no significant impact on the surface abundance of $^{4}$He since it develops only in the upper wing of the hydrogen-burning shell (see e.g. Paper I for more details)\footnote{Thermohaline mixing leads to an increase of the surface abundance of $^{4}$He during the RGB by $\sim$0.1\% in the 1.25~M$_{\odot}$ models independently of the stellar metallicity.}.
However, rotation-induced mixing smoothes the internal abundance gradients compared to the standard case, leading freshly produced $^{4}$He to diffuse outwards. This can be seen in Fig. \ref{fig:PXHe4} where we show the abundance profile of $^4$He at the end of the main sequence for two stellar masses and metallicities in the standard and the rotating cases (black full and red dashed lines, respectively; the vertical bars indicate the depth reached by the convective envelope at its maximum extent during the first dredge-up). This results in stronger $^{4}$He abundance variations at the end of the first dredge-up when rotation-induced mixing is accounted for, as seen in Fig.~\ref{fig:PXHe4_BS99} (bottom panels). For the same reasons, the effect of the second dredge-up is also strengthened in the rotating models. Overall, the impact of rotation increases with decreasing metallicity (see Paper I, and references therein). 

Consequently, the yields of $^{4}$He are higher in the models including rotation-induced mixing than in the standard case (see tables \ref{tableyields0001} to \ref{tableyields014}). And at the same time, the fraction of H is decreased.

Figure \ref{fig:Hrd_He4} summarizes all the effects described above on the surface abundance of $^{4}$He along the evolution of standard and rotating stars with different initial masses at solar metallicity. The tracks are shown from the zero age main sequence up to the end of the TP-AGB, with the colour coding given on the right scale.

\subsubsection{Comparison with other stellar models}
\label{4Hecomparison}

 In Fig.~\ref{fig:PXHe4_BS99} we compare our predictions with those of \citet{BoSa99}.
 We note a very good agreement as far as the standard models are concerned (upper panels). On the other hand, neither the thermohaline instability nor BS99 ``cool bottom processing'' do affect the $^4$He surface abundances. The differences that appear in the lower panels result essentially from the effects of rotation-induced mixing.

\section{Chemical evolution model}
\label{chemcode}

\subsection{Basic assumptions}

The adopted model for the chemical evolution of the Galaxy assumes that the 
Milky Way forms out of two main accretion episodes almost completely 
disentangled \citep{Chia97,Chia01}. During the first one, the primordial gas collapses very 
quickly and forms the spheroidal components, halo and bulge. During the second 
one, the thin-disc forms, mainly by accretion of matter of primordial 
chemical composition. The disc is built-up in the framework of the inside-out 
scenario of Galaxy formation, which ensures the formation of abundance 
gradients along the disc \citep{Larson76,MattFran89}. The 
Galactic disc is approximated by several independent rings, 2 kpc wide, without 
exchange of matter between them. 

\subsection{Nucleosynthesis prescriptions}
\label{sec:nucmod}

The nucleosynthesis prescriptions for metals (elements heavier than $^4$He) are 
from Paper~II for low- and intermediate-mass stars 
(M $\leq$ 6~M$_{\odot}$). As for massive stars and Type Ia supernovae, we adopt the same nucleosynthesis prescriptions 
as in \citet{RKTM10}, their model~6, namely:
\begin{itemize}
\item Yields for core-collapse supernovae are taken from \citet{Koba06}, except for carbon, nitrogen and oxygen, for which the adopted yields are from \citet{MeyMa02,Hirschi05,Hirschi07,Ekstroem08}.  
\item Yields for Type Ia supernovae are from \citet{Iwamoto99}.
\end{itemize}

As far as the nucleosynthesis prescriptions for D, $^3$He and $^4$He are concerned, we use standard prescriptions from \citet{Dearborn96} in the 6 to 100 M$_{\odot}$ mass range for D and $^3$He and yields by \citet{MeyMa02,Hirschi05,Hirschi07,Ekstroem08} for $^4$He. 
For low- and intermediate-mass stars (M $\leq$ 6~M$_{\odot}$) we use the yields from Papers II and III \footnote{$^3$He stellar yields are presented in Table 1 to 4 of Paper II. $^4$He stellar yields are shown in Tables \ref{tableyields0001} to \ref{tableyields014} } and we compute three different models with the following assumptions (summarized in Table \ref{tab:models}) : 

\begin{table}[t]
  \caption{Description of different models computed with yields from \citet{Lagarde11} for stars below 6M$_{\odot}$. }
    \centering
  \begin{threeparttable}
  \begin{tabular}{c c c }
    \hline
    Models &  \multicolumn{2}{c }{Stellar mass} \\ 
                  & M$\leq$ 2.5 M$_{\odot}$ & M$>$2.5 M$_{\odot}$\\  
    \hline
    \hline
    A 		& 100 \% standard		&  	100 \% standard	\\
    \hline
    B		& 96 \% th. + rot $^{[1]}$		&	100 \% th.+ rot	\\
    		& 4 \% standard		& 				\\
   \hline
    C		&	100\% th.+ rot		&	100 \% th.+ rot	\\
    \hline
  \end{tabular}
  \label{tab:models}
      \begin{tablenotes}
       \item [1] th. + rot means that we use yields including thermohaline instability and rotation induced mixing.
       
       	\end{tablenotes}
	\end{threeparttable}
\end{table}

\begin{itemize}
\item 
In Model A we consider the standard yields for all stars with masses below 6.0~M$_{\odot}$.
\item 
In Model B we consider the yields including thermohaline instability and rotation-induced mixing for 100\% of the intermediate-mass stars (M~$>$~2.5 M$_{\odot}$) and 96\% of the low-mass stars (M~$\leq$~2.5 M$_{\odot}$). The remaining 4\% of the low-mass stars are assumed to release the yields as predicted by the standard theory. 
This allows us to account for the fact that some rare planetary nebulae (J320, NGC 3242) exhibit $^{3}$He as predicted by the standard models (see e.g. \citealt{Balseretal2007}, and references therein).
\item In Model C we assume the yields including thermohaline instability and rotation-induced mixing for all the low- and intermediate-mass stars. Obviously the outcome of models B and C are expected to be very similar.
\end{itemize}

Linear interpolations in mass and metallicity are used to fill the gaps in the 
computed grids of yields (see \citealt{RKTM10} for a discussion of the potential spurious effects introduced by this procedure).

\begin{figure*} 
	\centering
		\includegraphics[angle=0,width=9cm]{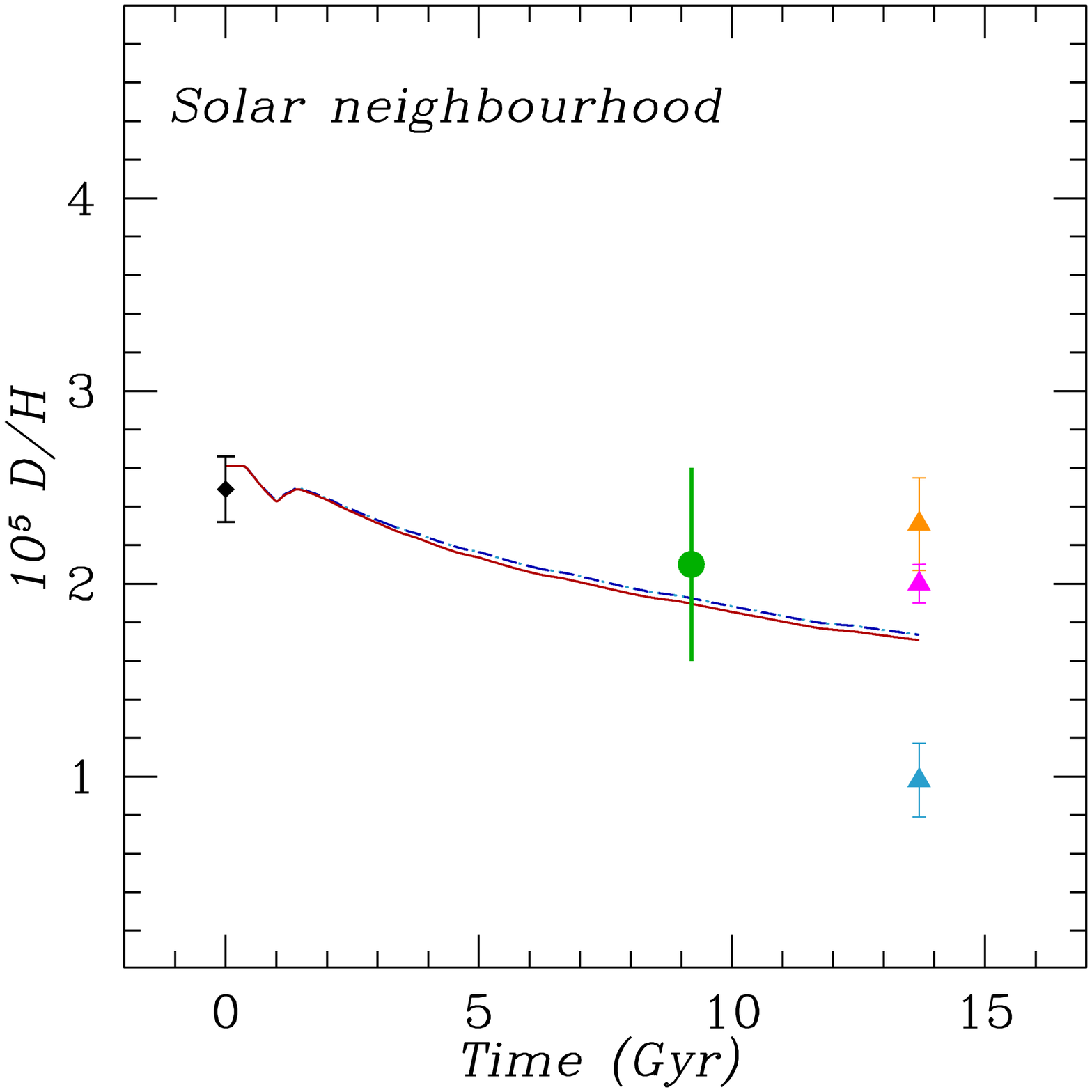}
		\includegraphics[angle=0,width=9cm]{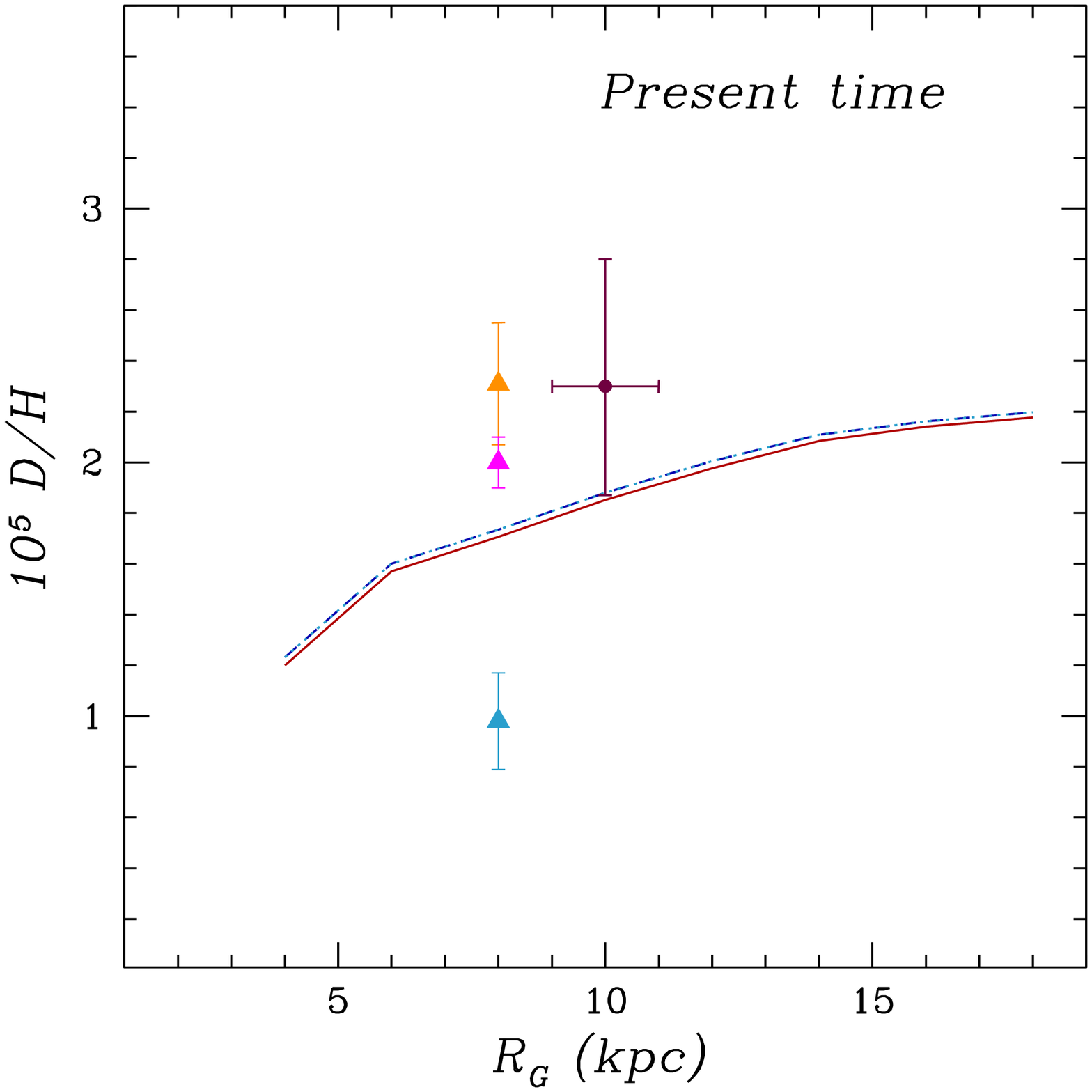}
	  \caption{Evolution of D/H in the solar neighborhood ({\textbf{\textit{left panel}}}) and distribution of D/H along the Galactic disc at the present time ({\textbf{\textit{right panel}}}). The predictions of Models A and B are shown (red solid and dashed blue lines, respectively). At t=0 Gyr, we plot the \emph{WMAP} value (filled black diamond). The PSC data by \citet{GeGl98} are shown (filled green circle). The local interstellar medium (LISM) data are from \citet{Linsky06}, \citet{Hebrard05} and \citet{Prodanovic10} (filled orange, light blue, and magenta triangles, respectively). The data for the outer disc (R$_{G}$=10 kpc) are from \citet[][filled bordeaux circle]{Rogers05}. }
	\label{fig:H2H}
\end{figure*}

\section{Evolution of the light elements in the Milky Way}
\label{lightelementevol}

In the following sections we discuss the evolution of the light elements in the Galaxy within the framework described in Section~\ref{chemcode}
when taking into account the yields from our non-standard stellar models, and compare the predictions with 
different observations. 
The relevant data are presented in Table~\ref{tab:obs}. 

\subsection{Evolution of deuterium} 

\begin{figure*} 
	\centering
		\includegraphics[angle=0,width=9cm]{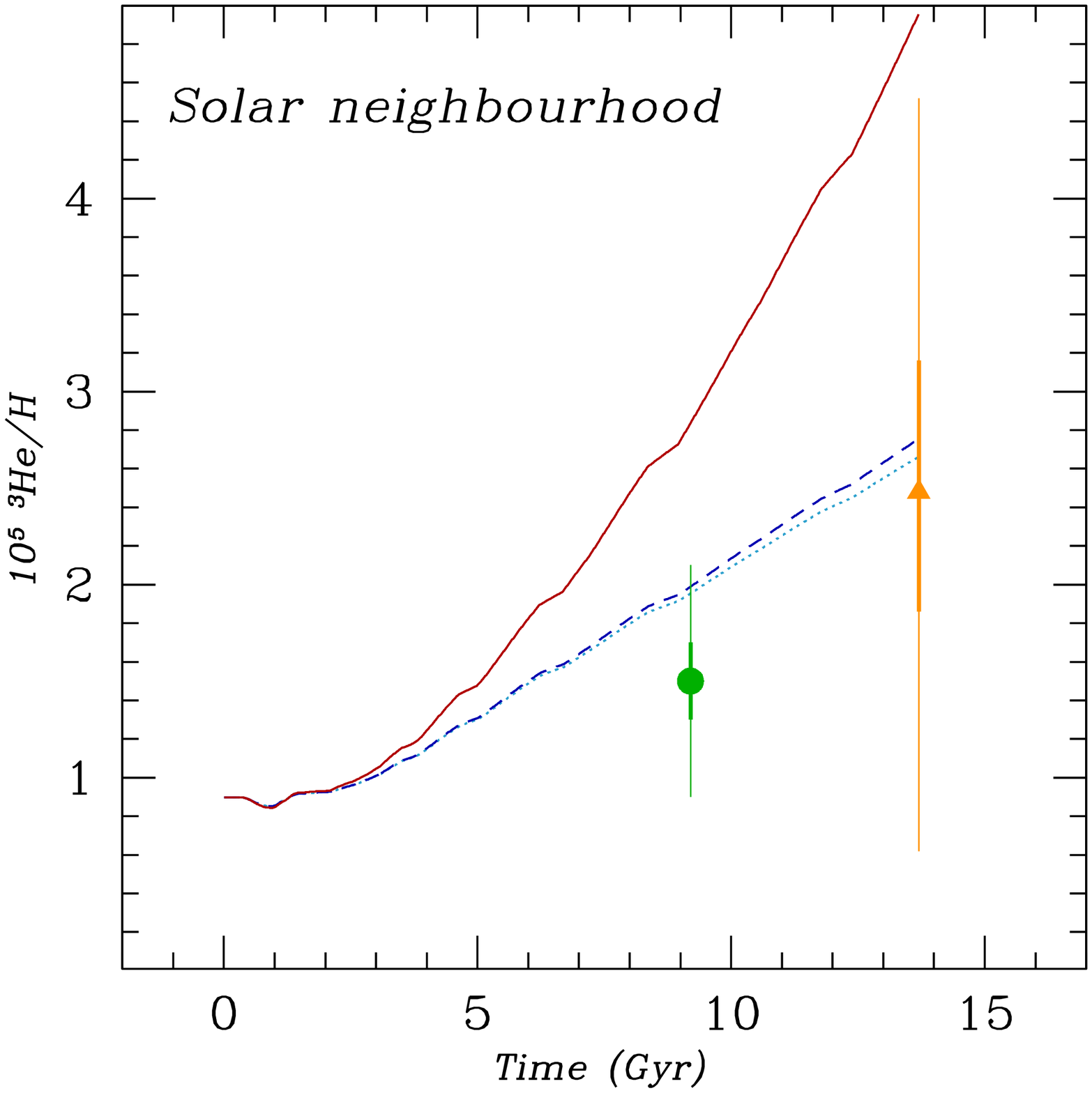}
		\includegraphics[angle=0,width=9cm]{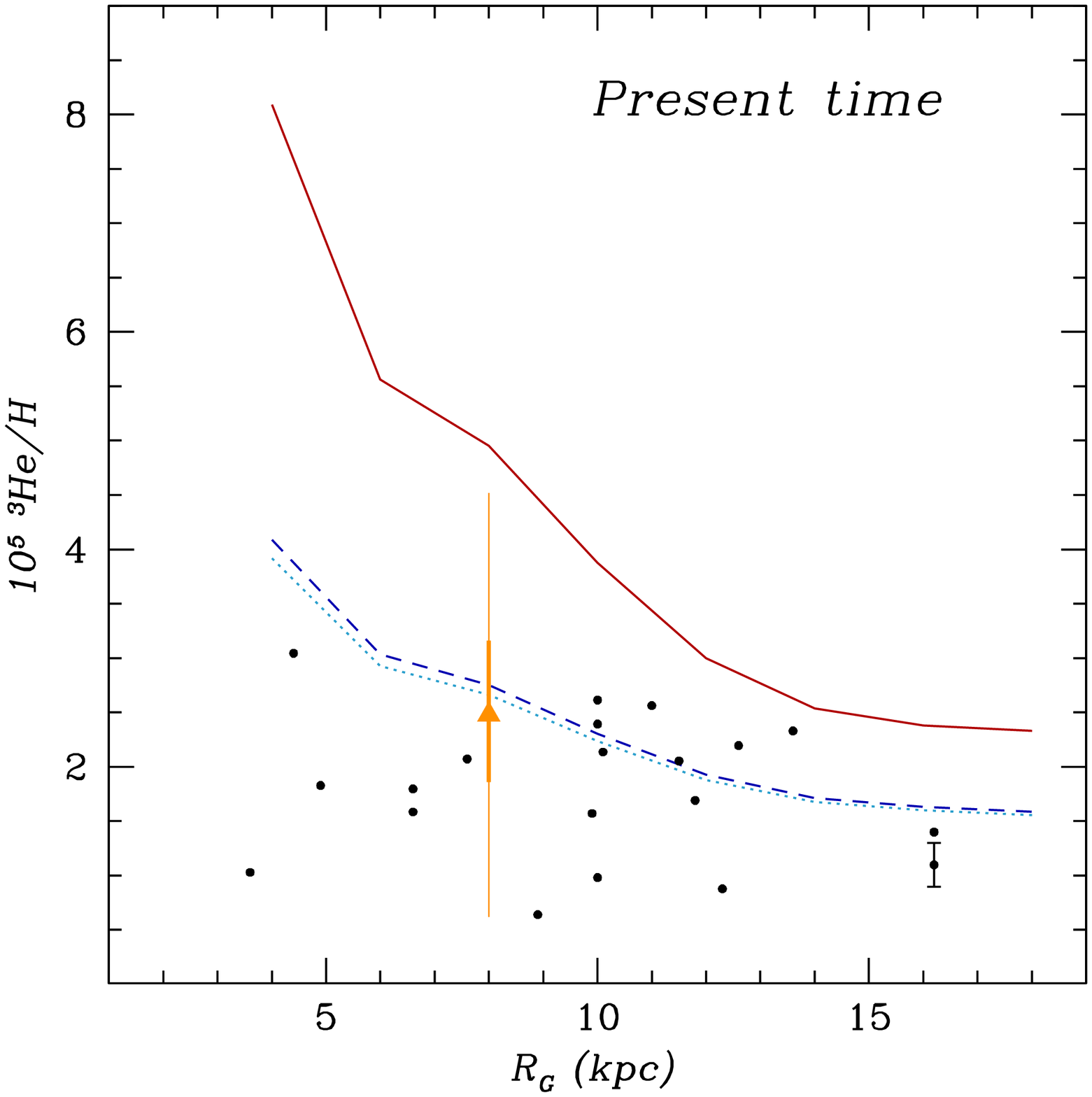}
	  \caption{\textbf{\textit{Left panel :}} Evolution of $^{3}$He/H with time in the solar neighborhood. Data for the PSC (green filled circle) and local interstellar medium (LISM, orange filled triangle) are from \citet{GeGl98} and \citet{GlGe96}, respectively. 1-$\sigma$ and 3-$\sigma$ error bars are shown with thick and thin lines, respectively. 
	  \textbf{\textit{Right panel :}} Radial distribution of $^{3}$He/H at the present time. The dots are HII regions data from \citet{Bania02} (error bars are shown only for S209; see text for discussion). The triangle at R$_{G}$=8 kpc represents LISM data from \citet{GlGe96}. The predictions from Models A, B and C are shown in both panels by the red full, blue dashed and cyan dotted lines respectively.}  
	\label{fig:He3}
\end{figure*}

In this paper we assume (D/H)$_{\mathrm{P}}$ = 2.6~$\times$ 10$^{-5}$ for consistency with the initial value adopted by the stellar models. However, we note that our GCE model gives a better
fit to the data when a value of (D/H)$_{\mathrm{P}}$ = 2.8~$\times$ 10$^{-5}$ as observed in quasar spectra \citep{Pettini08} is adopted instead \citep[see Fig.3 of ][]{Romano10}.

Figure \ref{fig:H2H} (left panel) shows the data for deuterium from the Big Bang (\emph{WMAP} value, filled black diamond) to the present day (local interstellar medium, LISM, filled triangles). 
The determinations of primordial and PSC (filled green circle, \citealt{GeGl98} according with \citealt{GeRe72, GeRe81}) deuterium abundances underline a small depletion from the Big Bang (t=0 Gyr) to the solar-system formation (t=9.2 Gyr). 
Analyses of \emph{Far Ultraviolet Spectroscopic Explorer (FUSE)} observations have allowed measurements of D/H in the LISM for many lines of sight. These observations have revealed a large variation of the local D abundances, which complicates the interpretation in the context of standard GCE models \citep{Linsky06,Linsky10}. \citet{Hebrard05} and \citet{Linsky06} proposed that either the lowest (D/H=0.98~$\times$ 10$^{-5}$) or the highest (D/H=2.31~$\times$ 10$^{-5}$) observed value is indicative of the true LISM value reflecting the process of D astration through successive stellar generations during the whole Galaxy's evolution. \citet{Hebrard05} suggest a value of the true local deuterium abundance lower than the one measured in the local bubble. On the other hand, \citet{Linsky06} give a lower bound to the true local deuterium abundance very close to the primordial abundance, pointing to a deuterium astration factor smaller than predicted by standard GCE models. More recently, \citet{Prodanovic10} have applyed a statistical Bayesian method to determine the true local D abundance. They propose a value very close to the D abundance at the time of the formation of the Sun (see Table \ref{tab:obs}). GCE models that fulfil all the major observational constraints available for the solar neighbourhood and for the Milky Way disc can explain the LISM D abundance suggested by \citet{Prodanovic10} as a result of D astration during Galactic evolution; lower and higher values can only be explained as due to small-scale, transient phenomena, such as D depletion on to dust grains and localized infall of gas of primordial chemical composition \citep[][and references therein]{Romano06,Steigmanetal07,Romano10}.
\\
\begin{figure*} 
	\centering
		\includegraphics[angle=0,width=9cm]{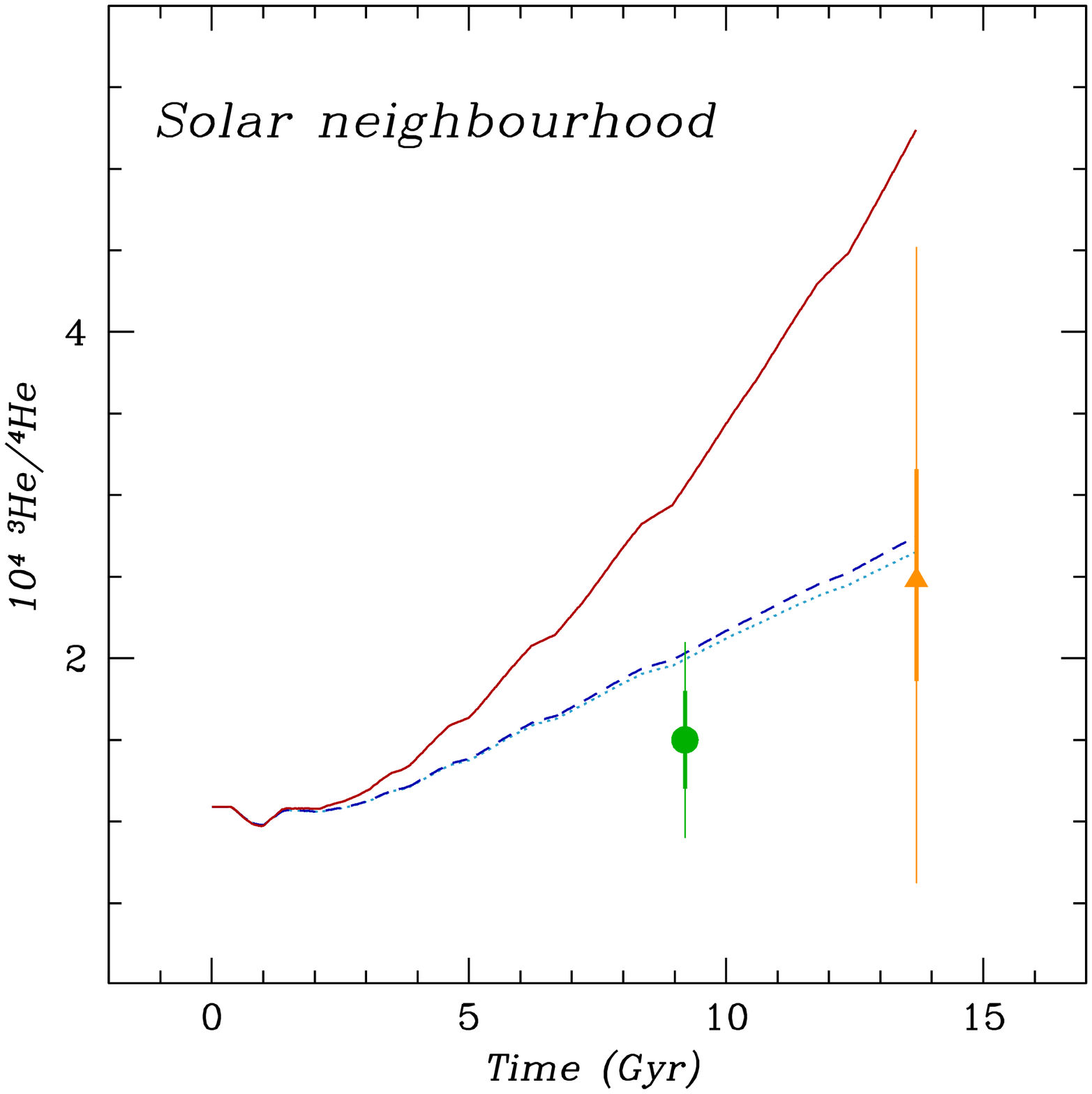}
		\includegraphics[angle=0,width=9cm]{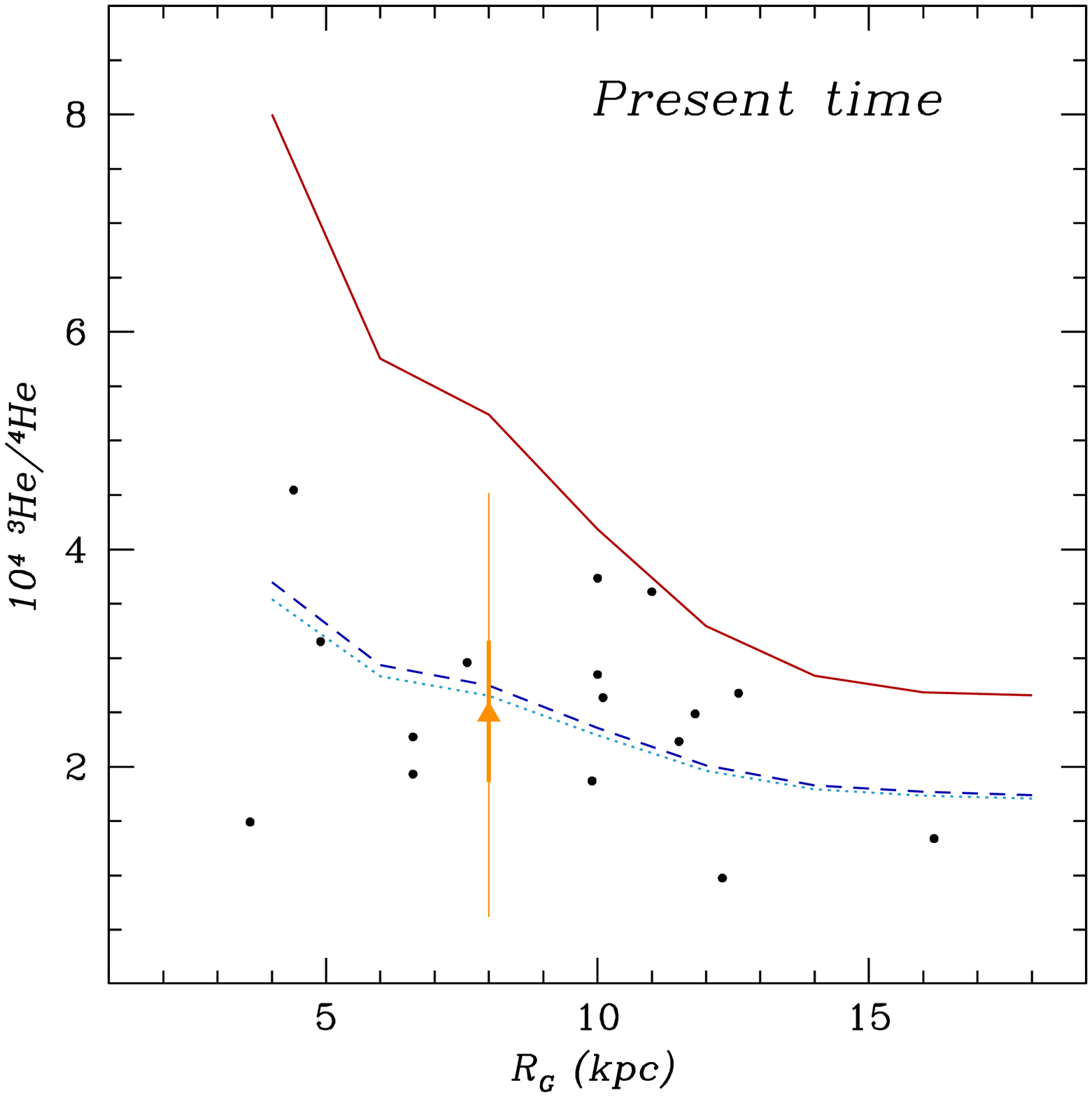}
	  \caption{Temporal (\textbf{\textit{left panel}}) and spatial (\textbf{\textit{right panel}}) variation of helium isotopic ratio in the solar neighborhood and along the galactic disk at the present time, respectively. Models are the same as in Fig.\ref{fig:He3}. Data are from \citet{GeGl98} for the PSC (green filled circle), \citet{Bania02} for Galactic HII regions (black dots), and \citet{GlGe96} for the LISM (orange filled trangle). 1-$\sigma$ and 3-$\sigma$ errors are also shown as thick and thin lines, respectively.}
	\label{fig:He3He4}
\end{figure*}

Our predictions for the evolution of deuterium with time in the solar neighborhood and for the present deuterium abundance profile along the Milky Way disc are shown in Fig.~\ref{fig:H2H} (left and right panels, respectively).
The very modest shift between the predictions of Models A and B is due to the slight difference in the H yields when stellar rotation is accounted for compared to the standard case. Since pristine deuterium burns entirely in stars independently of their mass and metallicity, the predicted trends depend only on the total astration assumed in the GCE model.
The present predictions fit well the PSC deuterium data and favour the local abundance of deuterium suggested by \citet{Prodanovic10}. 

\subsection{Evolution of $^{3}$He}

Figure \ref{fig:He3} (left panel) shows the evolution of $^{3}$He/H in the solar neighborhood as predicted by Models A, B, and C (see Sect.~\ref{sec:nucmod}).
As already discussed in the literature (see Sect.~\ref{section:introduction}), standard stellar models strongly overestimate the production of $^3$He by low-mass stars.

According to standard predictions, $^{3}$He is produced on the main sequence during the core hydrogen burning and it is not destroyed during later phases. Consequently, when adopting standard predictions $^{3}$He is overproduced in the course of Galactic evolution (Model A, red solid line). In low-mass stellar models including thermohaline mixing, $^{3}$He is destroyed from the bump luminosity on the RGB during shell hydrogen burning and during the early AGB. When low-mass stars are assumed to experience this extra-mixing, GCE models do not overproduce $^{3}$He (Fig.~\ref{fig:He3}, Model C, cyan dotted line).
Model B (dashed blue line) shows the effect of inhibiting the thermohaline mixing in 4\% of low-mass stars (see Section~\ref{sec:nucmod}), that thus follow the standard prescriptions. As can be seen in Fig.~\ref{fig:He3}, negligible differences are found between Model B and Model C predictions.
The PSC and local $^{3}$He abundances \citep{GeGl98,GlGe96} are reproduced by the models (at 3- and 1-sigma level, respectively).\\

The present-day $^{3}$He abundance distribution along the Galactic disk is shown in Fig.~\ref{fig:He3} (right panel). Model A, adopting standard $^3$He yields for low-mass stars, predicts too much $^{3}$He at all Galactocentric distances.  
In the inner regions, where the star formation was stronger, the contribution of low- and intermediate-mass stars to the $^{3}$He enrichment of the ISM is more important.  As a consequence, Model A predicts a large negative gradient of $^3$He/H. Although the contribution of low-mass stars in the inner regions is significantly reduced in Model B, it still predicts a negative gradient. In Fig.~\ref{fig:He3}, we compare the predictions of Models A and B with observations of $^3$He in HII regions from \citet{Bania02} and with the local $^{3}$He abundance from \citet{GlGe96}. Contrary to what is predicted by classical theory,  
the observations in HII regions show a gradient close to zero. \citet{Bania02} derived meaningful error estimates only for one source [S209 : $^{3}$He/H=(1.1$\pm$0.2)~$\times$ 10$^{-5}$) and pointed our the riskiness of basing one's conclusion on only one object. In order to better constrain both stellar and galactic evolution studies, it would be very useful if future work could provide a sound estimate of the errors associated to $^{3}$He determinations across the Galactic disk. Nevertheless, we can conclude that thermohaline and rotating mixing are required to fit currently available measurements of $^3$He in Galactic HII regions.
In addition, the $^{3}$He gradient is d\,log($^{3}$He/H)/d\,R$_{G}\sim-0.04$ and $-$0.028 dex/kpc for models A and B, respectively, consistent with that predicted by \citet{Chiappini02}, $-$0.04$<$d\,log($^{3}$He/H)/d\,R$_{G}$$<-$0.03 dex/kpc.

\citet{Tosi00} showed that GCE models including the CBP \citep{BoSa99} also fit the relevant Galactic $^3$He data, provided the primordial abundance of D is sufficiently low \citep[see also][]{Chiappini02,Romano03}. However, CBP destructs $^{3}$He too quickly and is not a physical mechanism (see Section \ref{He3_BS}). Therefore, it does not provide a reliable physical explanation of the $^{3}$He observations in HII regions.

\subsection{Evolution of $^{4}He$}

Figure \ref{fig:He3He4} shows the temporal (left panel) and the spatial (right panel) evolution of the helium isotopic ratio ($^{3}$He/$^{4}$He) in the solar neighborhood and along the Galactic disk at the present time, respectively. Only the predictions of Models A and B are shown in the right panel. The PSC and LISM data are from \citet{GeGl98} and \citet{GlGe96}, respectively. The standard theory predicts a significant evolution of the helium isotopic ratio in the Galaxy, and a strong gradient across the Galactic disc. When the thermohaline mixing -- which destroys $^{3}$He in the stellar interior -- and the rotation-induced mixing -- which increases $^{4}$He at the stellar surface (see Section 3.3) -- are taken into account, lower helium isotopic ratios are obtained (cfr. the dashed versus solid lines in Fig.~\ref{fig:He3He4}). Consequently, the gradient of the helium isotopic ratio across the Galactic disc decreases and the observations in HII regions can be reproduced.

\begin{figure} 
	\centering
		\includegraphics[angle=0,width=9cm]{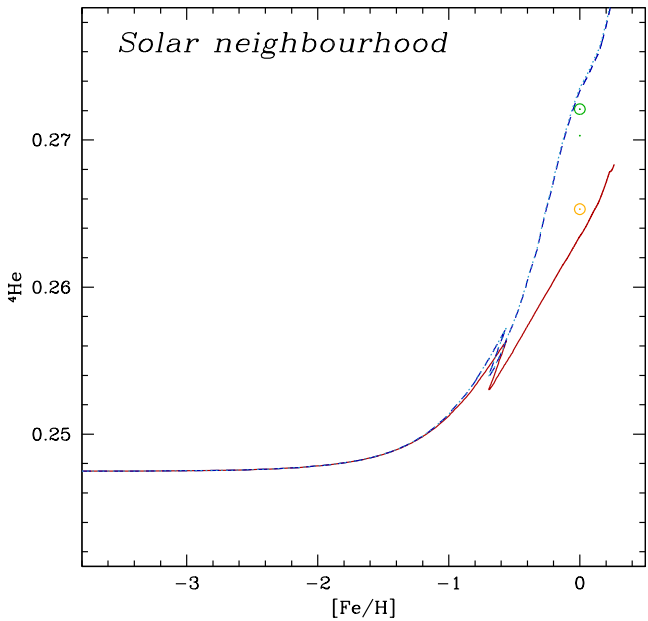}
	  \caption{Y versus [Fe/H] in the solar neighborhood predicted by models including yields computed with standard prescriptions (Model A; solid line) and with thermohaline mixing and rotation-induced mixing (Models B and C; dashed and dotted lines, respectively). Model predictions are compared with the protosolar values given by \citet{Serenelli10}.}
	\label{fig:He4}
\end{figure}

Figure \ref{fig:He4} presents the predictions for Y versus [Fe/H] in the solar neighborhood. The model predictions are compared with the initial abundance of $^{4}$He in the Sun derived by \citet{Serenelli10} from full standard solar models computed for different sets of solar abundances. When the solar abundances from \citet{GreSau98} and a metal to hydrogen ratio in the solar photosphere of ($Z/X$)$_{ph}$=0.0229 are adopted, a value of Y$_{i}$=0.2721 is obtained, which results in good agreement with the predictions of chemical evolution models including rotation-induced mixing and thermohaline mixing (green Sun symbol and dashed and dotted lines in Fig.\ref{fig:He4}, respectively). On the other hand, choosing the solar abundances from \citet{Asplund09} and ($Z/X$)$_{ph}$=0.0178 result in Y$_{i}$=0.2653, which is better fitted by models including standard stellar nucleosynthesis (orange Sun symbol and solid line in Fig.\ref{fig:He4}). Since the actual solar chemical composition is matter of debate, we can not discriminate between our models basing on the results for Y.

\section{Conclusions}
\label{conclusions}

In this article, we have described the results for the Galactic evolution of the primordial elements D, $^{3}$He and $^{4}$He in the light of new stellar models including the effects of thermohaline instability and rotation-induced mixing on the stellar structure, evolution and nucleosynthesis. The new stellar models are presented and discussed at length in \citet{ChaLag10} and \citet{Lagarde11}. We model the time-behavior of D, $^{3}$He and $^{4}$He in the solar neighbourhood, as well as in the inner and outer disc. The predictions of our models including thermohaline instability and rotation-induced mixing are compared to the predictions of models adopting standard nucleosynthesis prescriptions and to the most recent relevant observations. \\

We have seen that the CBP proposed before by \citet{BoSa99} is not a physical mechanism, and yields deduced do not take into account phases beyond the RGB.  
As discussed in the literature \citep{ChaZah07a,ChaLag10,Lagarde11}, thermohaline mixing can induces significant depletion of $^{3}$He in low- and intermediate-mass stars. Although these stars remain net producers of $^{3}$He, their contribution to the Galactic evolution of this element is highly reduced compared to classical theory. Indeed, our GCE models including thermohaline mixing reproduce the observations of $^3$He in the PSC, LISM and HII regions, while $^3$He is overproduced on a Galactic scale with standard models. Thermohaline mixing is the only  physical mechanism known so far able to solve the 
so-called ``$^{3}$He problem'' plaguing in the literature since many years. Importantly, its inhibition by a fossil magnetic field in red giant stars that are descendants of Ap stars does reconcile the measurements of $^3$He/H in Galactic HII regions with high values of $^3$He in a couple of planetary nebulae.\\
On the other hand, rotation has an impact on the stellar yields of H and $^{4}$He. Whether GCE models including rotation fit better the Galactic $^4$He data than standard models can not be said, because of the uncertainties on the actual solar chemical composition. However, they are consistent with the relatively high value of (D/H)$_{\text{LISM}}$ proposed by \citet{Prodanovic10}.

We conclude that GCE models including both thermohaline mixing and rotation-induced mixing reproduce satisfactorily well all the available data on D, $^3$He and, possibly, $^4$He abundances in the Milky Way, within the errors. However, the additional data on HII regions and better refining of their interpolation would be crucial to assess how good our fit actually is.  \\

\begin{table}
	\hspace{3cm}
	\caption{Model results for metallicity of Z=0.0001
	} 
	\scalebox{1.00}{

         \begin{threeparttable}
	\centering

	\begin{tabular}{| c | c | c | c || c |}  
	\hline  
	M     &                 & V$_{\rm ZAMS}$/V$_{crit}$ $^{[1]}$& V$_{\rm ZAMS}$$^{[1]}$ & Yield $^{4}$He $^{[2]}$ \\
	(M$_{\odot}$) & &                      &        (km.sec$^{-1}$)     & (M$_{\odot}$)              \\
	\hline \hline
	0.85 & stand.      & -  & -      & 5.22.10$^{-03}$    \\
		& th. +rot. & 0.45 &115 &  1.29.10$^{-02}$  \\
          \hline
	1.0   & stand.  & - & - & 9.41.10$^{-03}$   \\
	         & th. +rot.  & 0.45 &116  & 2.42 .10$^{-02}$ \\
          \hline
	1.25  & th. +rot.  & 0.45 &125 & 2.98.10$^{-02}$  \\
          \hline
	1.5   & stand.  & - & - & 2.44.10$^{-02}$  \\
	         & th. +rot.  & 0.45 &134 & 3.74 .10$^{-02}$   \\
          \hline
          2.0  & stand.     & - & -  & 3.38.10$^{-02}$   \\
                  & th. +rot.  & 0.45 &150  & 4.89.10$^{-02}$   \\
          \hline
         2.5   & stand.  & - &- & 1.90.10$^{-02}$  \\
                  & th. +rot.  & 0.45 & 162  & 2.77.10$^{-02}$ \\
          \hline
         3.0   & stand.  & - & - & 3.99.10$^{-02}$  \\
                  & th. +rot.  & 0.45 & 170  &   8.21.10$^{-02}$ \\
         \hline 
         4.0 & stand.  & - & - & 1.67.10$^{-02}$ \\                  
               & th. +rot.  & 0.45 & 152 & 2.75.10$^{-02}$  \\        
          \hline
         6.0  & stand.  & - & - & 4.09.10$^{-02}$  \\
                 & th. +rot.  & 0.45 & 175 & 4.79.10$^{-02}$ \\
         \hline
	\end{tabular}
	\label{tableyields0001}
	\begin{tablenotes}
        \item Each row contains entries for different assumptions: standard (without thermohaline or rotation-induced mixing); thermohaline mixing only; thermohaline and rotation-induced mixing. \\
        \item[1] The initial rotation on the ZAMS 
        \item[2] Yields of $^{4}$He
        
       	\end{tablenotes}
	
	\end{threeparttable}}

\end{table}

\begin{table}
	\hspace{3cm}
	\caption{Same as Table~\ref{tableyields0001} for Z=0.002 } 
	\scalebox{1.00}{ 
	\centering
	
	\begin{tabular}{| c | c | c | c || c |}  
	\hline  
	M     &                 & V$_{\rm ZAMS}$/V$_{crit}$ & V$_{\rm ZAMS}$ & Yield $^{4}$He  \\
	(M$_{\odot}$) & &                      &        (km.sec$^{-1}$)     & (M$_{\odot}$)              \\
	\hline \hline
	0.85 & stand.      & -  & -      &   7.47.10$^{-03}$ \\
		& th. +rot.  & 0.45 &114 &    1.61.10$^{-02}$ \\
          \hline
	1.0   & stand.  & - & - &   1.18.10$^{-02}$  \\
	         & th. +rot.  & 0.45 &112  &  2.68.10$^{-02}$  \\
          \hline
	1.25  & th. +rot.  & 0.45 &115 &  2.90.10$^{-02}$ \\
          \hline
	1.5   & stand.  & - & - &  3.13.10$^{-02}$  \\
	         & th. +rot.  & 0.45 &123 &  4.20 .10$^{-02}$    \\
          \hline
          2.0  & stand.     & - & -  &   2.13.10$^{-02}$  \\
                  & th. +rot.  & 0.45 &137  &   4.71.10$^{-02}$  \\
          \hline
         2.5   & stand.  & - &- &   2.15.10$^{-02}$  \\
                  & th. +rot.  & 0.45 & 146  &  5.84.10$^{-02}$  \\
          \hline
         3.0   & stand.  & - & - &   9.84.10$^{-03}$ \\
                  & th. +rot.  & 0.45 & 153  &   5.76.10$^{-02}$  \\
         \hline 
         4.0 & stand.  & - & - &  1.28.10$^{-01}$  \\                  
               & th. +rot.  & 0.45 & 163 &  2.28.10$^{-01}$  \\        
          \hline
         6.0  & stand.  & - & - &   3.92.10$^{-01}$  \\
                 & th. +rot.  & 0.45 & 170 &  4.30.10$^{-01}$  \\
         \hline
	\end{tabular}}
	\label{tableyields002}

\end{table}

\begin{table}
	\hspace{3cm}
	\caption{Same as Table~\ref{tableyields0001} for Z=0.004 	} 
	\scalebox{1.00}{ 
	\centering
	\begin{tabular}{| c | c | c | c || c |}  
	\hline  
	M     &                 & V$_{\rm ZAMS}$/V$_{crit}$ & V$_{\rm ZAMS}$ & Yield $^{4}$He  \\
	(M$_{\odot}$) & &                      &        (km.sec$^{-1}$)     & (M$_{\odot}$)              \\
	\hline \hline
	1.0   & stand.  & - & - &  1.24.10$^{-02}$ \\
	         & th. +rot.  & 0.45 &112  &  2.10.10$^{-02}$\\
          \hline
	1.25  & th. +rot.  & 0.45 &111 & 2.84.10$^{-02}$\\
          \hline
	1.5   & stand.  & - & - & 1.95.10$^{-02}$ \\
	         & th. +rot.  & 0.45 &119 &  2.83.10$^{-02}$  \\
          \hline
          2.0  & stand.     & - & -  & 1.90.10$^{-02}$  \\
                  & th. +rot.  & 0.45 &123  & 3.33.10$^{-02}$  \\
          \hline
         2.5   & stand.  & - &- & 2.94.10$^{-02}$  \\
                  & th. +rot.  & 0.45 & 141  & 5.62.10$^{-02}$\\
          \hline
         3.0   & stand.  & - & - & 2.03.10$^{-02}$ \\
                  & th. +rot.  & 0.45 & 147  &  6.75.10$^{-02}$ \\
         \hline 
         4.0 & stand.  & - & - &7.83.10$^{-02}$ \\                  
               & th. +rot.  & 0.45 & 147 & 1.93.10$^{-01}$\\        
          \hline
         6.0  & stand.  & - & - & 3.49.10$^{-01}$ \\
                 & th. +rot.  & 0.45 & 167 & 4.14.10$^{-01}$\\
         \hline
	\end{tabular}}
	\label{tableyields004}

\end{table}

\begin{table}
	\hspace{3cm}
	\caption{Same as Table~\ref{tableyields0001} for Z=0.014 
	} 
	\scalebox{1.00}{ 
	\centering

	\begin{tabular}{| c | c | c | c || c |}  
	\hline  
	M     &                 & V$_{\rm ZAMS}$/V$_{crit}$ & V$_{\rm ZAMS}$ & Yield $^{4}$He  \\
	(M$_{\odot}$) & &                      &        (km.sec$^{-1}$)     & (M$_{\odot}$)              \\
	\hline \hline
	1.0   & stand.  & - & - &  1.29.10$^{-02}$ \\
          \hline
	1.25  & th. +rot.  & 0.45 &110 & 2.27.10$^{-02}$\\
          \hline
	1.5   & stand.  & - & - & 2.79 .10$^{-02}$\\
	         & th. +rot.  & 0.45 &110 & 2.45 .10$^{-02}$  \\
          \hline
          2.0  & stand.     & - & -  & 1.65.10$^{-02}$ \\
                  & th. +rot.  & 0.45 &110  &  3.36.10$^{-02}$ \\
          \hline
         2.5   & stand.  & - &- & 2.35.10$^{-02}$  \\
                  & th. +rot.  & 0.45 & 130  & 5.36.10$^{-02}$\\
          \hline
         3.0   & stand.  & - & - &  4.08.10$^{-02}$\\
                  & th. +rot.  & 0.45 & 136  & 8.09.10$^{-02}$  \\
         \hline 
         4.0 & stand.  & - & - & 4.80.10$^{-02}$ \\                  
               & th. +rot.  & 0.45 & 144 & 8.98.10$^{-02}$ \\        
          \hline
         6.0  & stand.  & - & - & 3.07.10$^{-01}$ \\
                 & th. +rot.  & 0.45 & 156 & 3.84.10$^{-01}$\\
         \hline
	\end{tabular}}
	\label{tableyields014}

\end{table}

\begin{acknowledgements}
We dedicate this paper to Robert T. Rood : the scientist to whom the stellar evolution and $^{3}$He communities owe so much, and the friend whom we will miss for ever.
We thank T. Bania, D.Balser, and H.Reeves for fruitful discussions. 
We are thankful to our referee, Dr Achim Weiss, for his interesting and constructive remarks on our article.
C.C., D.R., and M.T. gratefully acknowledge the enlightening and fascinating conversations with Johannes Geiss and his hospitality at the International Space Science Institute (ISSI) in Bern (CH).
We acknowledge financial support from the Swiss National Science Foundation (FNS) and the french Programme National de Physique Stellaire (PNPS) of CNRS/INSU.
\end{acknowledgements}

\bibliographystyle{aa}
\bibliography{Reference_REV}

\end{document}